\documentclass[pdftex]{pasj01}

\begin{document} 
\Received{2019/02/06}
\Accepted{2019/06/05}

\title{Near-Infrared Imaging Polarimetry toward  M17 SWex}

\author{Koji \textsc{Sugitani}\altaffilmark{1}%
}
\altaffiltext{1}{Graduate School of Natural Sciences, 
Nagoya City University, Mizuho-ku, Nagoya 467-8501, Japan}
\email{sugitani@nsc.nagoya-cu.ac.jp}

\author{Fumitaka \textsc{Nakamura}\altaffilmark{2}}
\altaffiltext{2}{National Astronomical Observatory, Mitaka, Tokyo 181-8588, Japan}

\author{Tomomi \textsc{Shimoikura}\altaffilmark{3}}
\altaffiltext{3}{Department of Astronomy and Earth Science, Tokyo Gakugei University, Koganei, Tokyo 184-8501, Japan}

\author{Kazuhito \textsc{Dobashi}\altaffilmark{3}}

\author{Quang \textsc{Nguyen-Luong}\altaffilmark{4,1}}
\altaffiltext{4}{IBM Canada, 120 Bloor Street East, Toronto, ON, M4Y 1B7, Canada}

\author{Takayoshi \textsc{Kusune}\altaffilmark{2}}

\author{Takahiro \textsc{Nagayama}\altaffilmark{5}}
\altaffiltext{5}{Kagoshima University, 1-21-35 Korimoto, Kagoshima 890-0065, Japan}

\author{Makoto \textsc{Watanabe}\altaffilmark{6}}
\altaffiltext{6}{Department of Applied Physics, Okayama University of Science, 1-1 Ridai-cho, Kita-ku, Okayama 700-0005, Japan}

\author{Shogo \textsc{Nishiyama}\altaffilmark{7}}
\altaffiltext{7}{Miyagi University of Education, 149 Aramaki-aza-Aoba, Aobaku, Sendai, Miyagi 980-0845, Japan}

\author{Motohide \textsc{Tamura}\altaffilmark{8,9,2}}
\altaffiltext{8}{ Department of Astronomy, The University of Tokyo, 7-3-1, Hongo, Bunkyo-ku, Tokyo, 113-0033, Japan}
\altaffiltext{9}{ Astrobiology Center of NINS, 2-21-1, Osawa, Mitaka, Tokyo 181-8588, Japan}



\maketitle

\begin{abstract}
We conducted  near-infrared ($JHK$s) imaging polarimetry 
toward the infrared dark cloud (IRDC) M17 SWex, including almost all of the IRDC filaments 
as well as its outskirts, with the polarimeter SIRPOL on the IRSF 1.4~m telescope.
We revealed the magnetic fields of M17 SWex with our polarization-detected sources 
that were selected by some criteria based on their near-IR colors and the column densities 
toward them, which were derived from the Herschel data.
The selected sources indicate not only that the ordered magnetic field is 
perpendicular to the cloud elongation as a whole, 
but also that at both ends of the elongated cloud the magnetic field appears to bent toward its central part, 
i.e., large-scale hourglass-shaped magnetic field perpendicular to the cloud elongation.
In addition to this general trend, the elongations of the filamentary subregions
within the dense parts of the cloud appear to be mostly perpendicular to their local magnetic fields, 
while the magnetic fields of the outskirts appear to follow the thin filaments that protrude from the dense parts.
The magnetic strengths were estimated to be $\sim$70--300  $\mu$G in the subregions, 
of which lengths and average number densities are $\sim$3--9 pc
 and $\sim$2--7$\times10^3$ cm$^{-3}$, respectively, 
by the Davis-Chandrasekhar-Fermi method with the angular dispersion of our polarization data and 
the velocity dispersion derived from the C$^{18}$O ($J$ = 1-0) data obtained by the Nobeyama 45~m telescope. 
These field configurations and our magnetic stability analysis of the subregions imply 
that the magnetic field have controlled the formation/evolution of the M17 SWex cloud. 
\end{abstract}

\section{Introduction}

Star formation rate in the Galaxy is known to be low (e.g., \cite{zuckerman1974, krumholz2007}) and 
the cause of this low rate is still open to debate.  
Three main processes, magnetic field, stellar feedback, and cloud turbulence, 
are thought as the agents
that regulate and control star formation against gravity in molecular clouds. 
Although not a little theoretical works have been done  (e.g., \cite{shu1987,mckee2007,krumholz2014}), 
it is not always clear which process is the major cause of the low rate 
or what combination among them is important to regulate star formation.  
Among these processes, candidates for stellar feedback and cloud turbulence have been extensively investigated 
by Mid-IR and/or radio surveys by covering the entire molecular clouds (e.g., \cite{churchwell2006,urquhart2007,arce2010,nakamura2011,narayanan2012}). 
However, the magnetic fields of individual, entire molecular clouds had not always been well investigated due to sparse 
and/or shallow sampling data of polarization (e.g., Taurus: \cite{heyer1987}, \cite{tamura1987}; 
Ophiuchus: \cite{wilking1979}; Several dark clouds: \cite{vrba1976}). 
To understand well  the role of the magnetic field in molecular clouds, 
it is essential to know their entire field structures with higher resolution, including those in their rather dense parts.  
Only in the 2010s, the magnetic field structures of molecular clouds have been 
investigated extensively and in detail  
(e.g., Taurus: \cite{chapman2011}; Ophiuchus: \cite{kwon2015}; Vela C: \cite{kusune2016}, \cite{soler2017a};  
Several nearby clouds: \cite{planckXXXV}), but these studies have not always been done in the context of star formation.   
\citet{marchwinski2012} examined the field structure and magnetic stability in the entirety of 
a quiescent molecular cloud GRSMC 45.60+0.3 ($d\sim1.9$ kpc) with their near-IR data (GPIPS: \cite{clemens2012}) 
and implied that the magnetic fields suppress star formation.  
However, the polarization studies that cover entire molecular clouds with reasonably high resolution 
probing local magnetic fields as well is still not many, particularly for a bit distant molecular clouds.
For example, SCUBA2-POL large programs BISTRO and BISTRO2 mapped numerous star-forming clouds 
in submm-polarimetric mode, but only focus on high density regions \citep{ward-thompson2016,pattle2017}.
Note that the polarized submm emission arises from the asymmetric dust grains aligned with orthogonal 
to the magnetic field and can be well detectable toward the high density regions of strong submm emission, 
while starlight polarization in the optical and near-IR arises from the dichroic extinction due to the aligned grains.  
These account for the polarization directions; the submm polarization is orthogonal to the magnetic filed, while the starlight polarization is parallel. 
Also note that the submm and starlight polarizations sample the plane-of-sky component of the magnetic field 
integrated along the line of sight.  
The recent understanding of dust grain alignment is reviewed in  \citet{andersson2015}. 

\begin{figure}
   \FigureFile(80mm,*){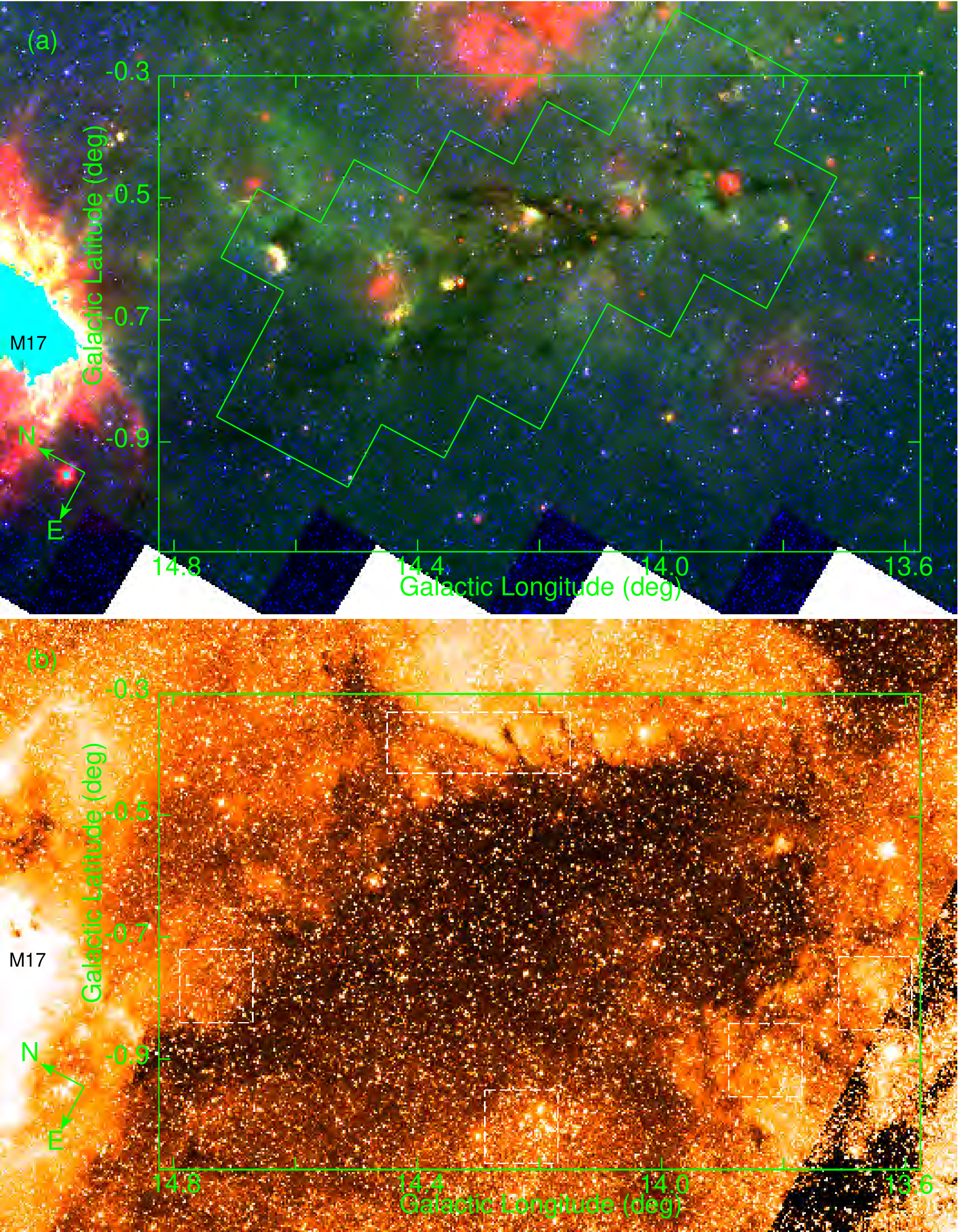} \\
\caption{(a) Three-color image of M17 SWex, which is produced from the Spitzer archival data 
(Red; 24 $\mu$m, Green; 8 $\mu$m, Blue: 3.5 $\mu$m).
An area of our near-IR imaging polarimetry toward M17 SWex is show by a green polygon.
A green galactic coordinate box corresponds to the area 
where we made a column density map from the Herschel archival image data.
A part of M17 (M17SW) is seen at the left side of this image.
(b) DSS2 R-band image.  This optical image is emphasized by histogram equalization method 
to clearly show the entire cloud of M17 SWex, including its outskirt part of lower column density, 
which is not clearly indicated in the mid-infrared image above.  
The reference areas that were used to subtract the background emission to construct the column 
density mapof the M17 SWex cloud are shown by dashed white boxes.}\label{fig1}
\end{figure}

The M17 SWex cloud \citep{povich2010} is one of the most remarkable, filamentary infrared dark clouds (IRDCs) 
with a distance relatively close to the Sun (figure \ref{fig1}a). 
Here, we adopt 2.0 kpc as a distance of this cloud, following \citet{povich2016}. 
The M17 SWex cloud seems to belong to the G16.8-M16/M17 molecular cloud complex \citep{nguyen2016}.
\citet{elmegreen1979} first found this cloud as one that extends for more than $1\degree$ in the southwest of M17  
(figure \ref{fig1}b) and estimated its gas mass to be $2\times10^5$ $M_\odot$ with $V_{\rm LSR}=20\pm2$ km s$^{-1}$, 
suggesting that it is the second generation OB star formation site next to M17.  

\citet{povich2010} searched this region for young stellar objects via IR excess emission 
and identified many intermediate-mass ones, of which the number is comparable to 
that of the central NGC 6618 cluster of M17, but not high-mass ones, 
i.e., their observed YSO mass function is significantly steeper than the Salpeter IMF 
on the higher-mass side. 
\citet{povich2016} further searched for more member stars of the M17 SWex cloud, 
including pre-main-sequence stars that lack IR excess emission from circumstellar disks in X-ray 
and confirmed that the lack of O-type stars is real. 
Based on the presence of many, outspread intermediate-mass pre-main-sequence stars without inner dust disks, 
they suggested that the M17 SWex cloud has been an active star-forming site for more than 1 Myr, 
but lacks high-mass stars.
They concluded, from these results, that M17 SWex is either an example that produces rich intermediate-mass stars, 
but few high-mass stars, or an example that have high-mass cores that are still accreting mass enough to form high-mass stars. 

Near-IR and optical observations were made toward the M17 SWex cloud \citep{busquet2013,santos2016}. 
However, their observed areas at near-IR are limited only toward the two dense hubs (Hub-N and Hub-S), 
which were studied by interferometric observations \citep{busquet2013, busquet2016}. 
In this paper, we present a more complete polarimetric observation of the entire M17 SWex cloud at $J$, $H$, $K$s, 
using the polarimeter SIRPOL on the IRSF 1.4~m telescope together with $^{13}$CO (J=1-0) 
and C$^{18}$O (J=1-0) data  obtained in the Nobeyama Radio Observatory (NRO) 45~m legacy project 
"Nobeyama 45~m mapping observations toward nearby clouds" \citep{nakamura2019b}.  
The detailed observational results for the individual regions using molecular line data are given in other articles 
(Orion A: \cite{tanabe2019}, \cite{ishii2019}, \cite{nakamura2019a}; 
Aquila Rift: \cite{shimoikura2019a}, \cite{kusune2019}; M17: \cite{shimoikura2019b}; \cite{nguyen2019}; 
other regions: \cite{dobashi2019a}, \cite{dobashi2019b}). 
After describing our polarimetric observations shortly in section 2, we describe our treatments of the archival data 
and the reduction and selection of our near-IR polarimetry data for analysis in section 3. 
The polarimetry results and comparison with the $^{13}$CO and C$^{18}$O data are shown in section 4.
After deriving magnetic field parameters, we discuss the magnetic structure and stability of the cloud in section 5. 

\section{Observations}

Simultaneous $JHK$s polarimetric observations were carried out toward the M17 SWex cloud  
in August 2012, March 2013, July—August 2013,  and July 2016.
The observed area is indicated by a polygon in figure \ref{fig1}a.
We used the imaging polarimeter SIRPOL, polarimetry mode of the SIRIUS camera \citep{kandori2006}, 
mounted on the IRSF 1.4 m telescope at the South African Astronomical Observatory. 
The SIRIUS camera is equipped with three 1024 $\times$ 1024 HgCdTe (HAWAII) arrays, 
$JHK$s filters, and dichroic mirrors, which enables simultaneous $JHK$s observations 
\citep{nagashima99,nagayama03}.
The field of view at each band is $7'.7 \times 7'.7$ with a pixel scale of 0$\arcsec$.45, and  27 fields were observed in total.

We obtained 10 dithered exposures, each 15 sec. long, at four wave-plate angles 
($0\degree$, $22\degree.5$, $45\degree$, and $67\degree.5$ in the instrumental coordinate system) 
as one set of observations and repeated this six times. 
The total on-target exposure time for each field was 900 sec. per wave-plate angle.  
Twilight flat-field images were obtained at the beginning and/or end of the observations. 
The average seeing was $\sim$1.\arcsec3 at $H$ 
during the observations with a range of $\sim$0.\arcsec8—2.\arcsec0.  

\section{Archival data and Data Reduction}

\subsection{Herschel Data}

We used the Herschel Science Archival SPIRE/PACS image data (Quality: level 2.5 processed)
\footnote{Maps from the combined scan and cross-scan observation for science analysis.
See the details in the documents "SPIRE Products Explained" and "PACS Products Explained" 
(https://www.cosmos.esa.int/documents/12133/1035800/SPIRE+Products\\+Explained 
and https://www.cosmos.esa.int/documents/12133/996891/PACS\\+Products+Explained, respectively).} 
to derive the column density map of M17 SWex. 

First, We convolved the 160, 250, and 350 $\mu$m images to the 500 $\mu$m resolution of 36\arcsec  
~by using the IDL package developed by \citet{aniano2011}.  
Then, we resampled up or down all the images (other than the 250 $\mu$m) to 
the same grid size of the  250 $\mu$m image (6\arcsec).
In order to isolate dust emission from M17 SWex, we subtracted the background, 
diffuse emission of the Galactic plane by 2D linear-plane fitting of several surrounding 
reference regions, which are shown by white dashed boxes (see figure \ref{fig1}b). 
We chose these regions in the places just outside the dark area seen in the optical.
After subtracting the background emission from each image, 
we derived spectral energy distribution (SED) at each position of the 250 $\mu$m image 
and calculated dust temperature ($T_{\rm D}$) and column density ($\Sigma$) 
in a similar way to \citet{konyves2010}.  
Assuming single temperature of the dust emission, a gray-body SED fitting was performed 
with a function of $I_\nu = B_\nu (T_D) (1-e^{\tau_\nu})$, 
where $B_\nu$ expresses the Planck law, $I_\nu$ is the observed surface brightness at frequency $\nu$, 
$\tau_\nu =\kappa_\nu \Sigma$ is the dust opacity, and $\kappa_\nu$ is the dust opacity per unit mass.  
We adopted $\kappa_\nu = 0.1 (\nu / 1000 {\rm GHz})^\beta$ cm$^2$/g and $\beta$ = 2, following \citet{konyves2010}.
They mentioned that the uncertainty of the mass estimate is a factor of $\sim2$ mainly 
due to the uncertainties in the dust opacity ($\kappa_\nu$).

We  excluded the points where the dust emission was detected below 3 times the rms noise, 
which was measured in the reference areas, in any band. 
In  the case that SED fitting failed or the surface brightness is not high enough for SED fitting, 
we set the pixel to NaN. 
As the data weight for the SED fitting, we adopted 1$/\sigma^2$, where $\sigma^2$ is 
the square sum of the rms noise 
and the calibration uncertainties of surface brightnesses (15\% at 500/350/250 $\mu$m from \cite{griffin2010}; 
20\% at 160 $\mu$m  from \cite{poglitsch2010}).    

\begin{figure}
 \begin{center}
 \FigureFile(80mm,*){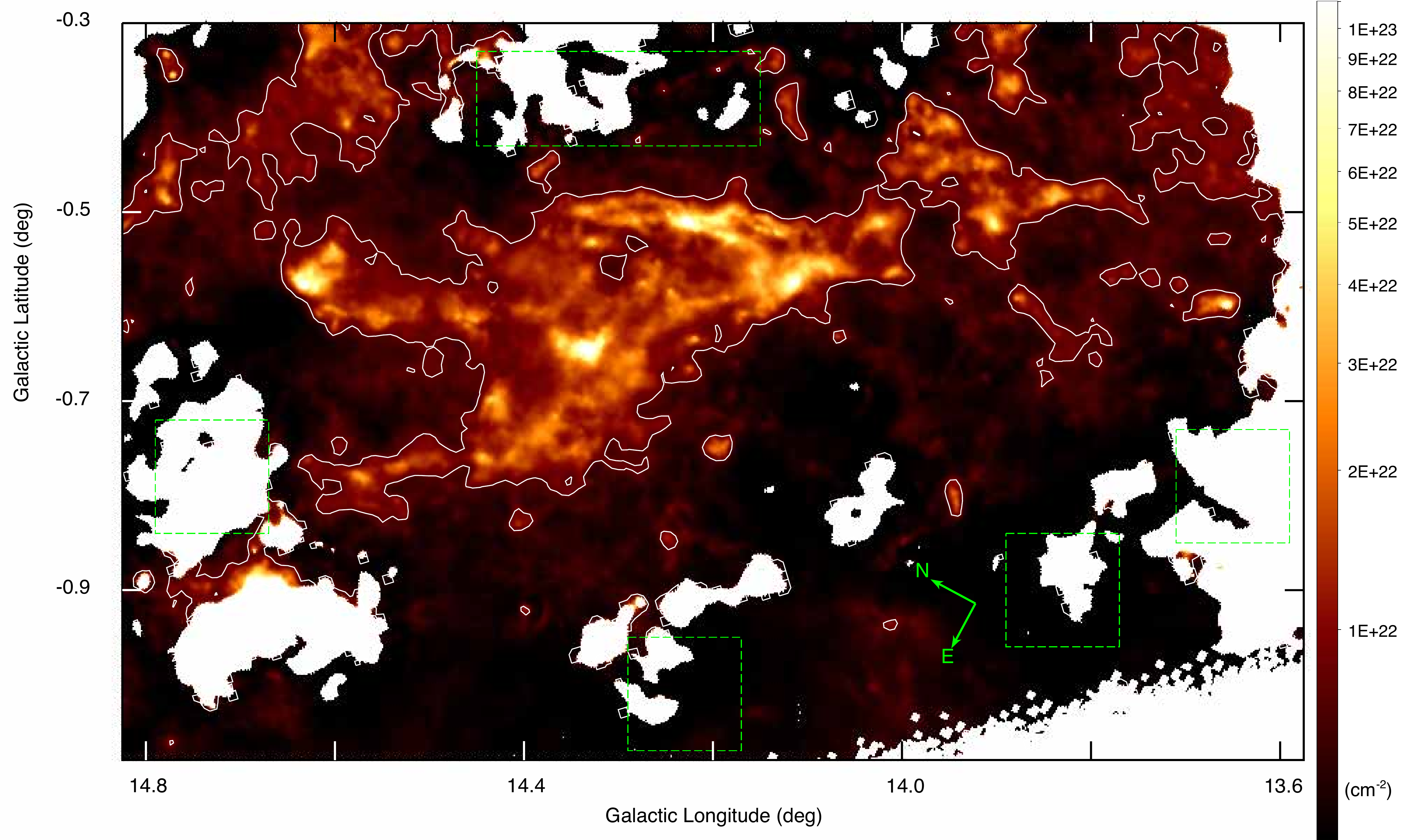} 
 \end{center}
\caption{H$_2$ column density map   of the M17 SWex cloud, 
which was derived by SED fitting of the Herschel archival data (160, 250, 350, and 500 $\mu$m).  
The reference areas that were used to subtract the background emission of the galactic plane
to construct the column density map are shown by five dashed green boxes.  
The subtraction was simply performed by 2D linear-plain fitting using the flux of these five areas 
to exclude the background emission from the galactic plane.  
The contour lines of $7.0\times10^{21}$cm$^{-2}$ are shown by white, thin lines.}\label{fig2}
\end{figure}

For better identification with the filaments in the Mid-IR image (figure \ref{fig1}a) and 
in the C$^{18}$O channel map that is presented in the next subsection, 
we obtained a column density map directly from the 250 $\mu$m surface brightness with a resolution of 18\arcsec, 
adopting the dust temperature obtained by the SED fitting of a lower resolution of 36\arcsec   
~and the mean molecular weight of 2.8. 
Here, we assume that the dust temperature gradually changes.  
The obtained column density map is shown in figure \ref{fig2}. 
In figure \ref{fig2}, we show the contour lines of $N_{{\rm H}_2}=7.0\times10^{21}$cm$^{-2} (A$v$\sim7)$, 
which was suggested to be a column density threshold for prestellar core formation \citep{konyves2015}. 

\subsection{$^{13}$CO ($J$=1-0) and C$^{18}$O ($J$=1-0) data}

We used the $^{13}$CO ($J$=1-0) and C$^{18}$O ($J$=1-0) data cubes of  
the NRO legacy project "Nobeyama 45~m mapping observations toward nearby clouds" 
using the NRO 45~m telescope. 
In these data, the effective angular resolution is 22\arcsec—24\arcsec ~and the grid interval is 7.5\arcsec.
\citet{shimoikura2019b} and \citet{nguyen2019} report the global cloud kinematics and dense core analysis 
with these data including the $^{12}$CO ($J$=1-0), N$_2$H$^+$ ($J$=1-0) and 
CCS ($J_N=8_7-7_6$) data, and discuss the detailed cloud structure and kinematics.

\subsection{Near-IR Polarization Data and Reduction}

To construct the $J$, $H$, $K$s images for polarimetry, we use a pipeline software 
of pyIRSF (https://sourceforge.net/projects/irsfsoftware/).  
This software executes standard procedures of dark subtraction, flat-fielding with twilight-flats, 
bad-pixel substitution, self-sky subtraction, and averaging of dithered images. 

Aperture polarimetry was performed each at $J$, $H$, and $K$s 
with an aperture radius of $\sim1$~FWHM by using APHOT of the DAOPHOT package. 
We calculated the Stokes parameters as follows: $I = (I_0+I_{22.5}+I_{45}+I_{67.5})/2$, 
$q = (I_0-I_{45})/I$, and $u = (I_{22.5} - I_{67.5})/I$, where $I_0$, $I_{22.5}$, $I_{45}$, and $I_{67.5}$
are intensities at four wave-plate angles. 
To obtain the Stokes parameters in the equatorial coordinate system ($q'$, and $u'$), 
a rotation of 105$^\circ$ \citep{kandori2006, kusune2015} was applied to them. 
The absolute accuracy of the position angle (P.A.) of polarization was estimated to be 
3$^\circ$ or less \citep{kandori2006, kusune2015}.
We calculated the degree of polarization $P$, and the polarization angle $\theta$ 
in the equatorial coordinate system as follows: $P= \sqrt{q^2 + u^2}$,  
$\theta = (1/2)$tan$^{-1}(u'/q')$. 
The foreground corrections in $q' $and $u'$ were not applied here because of 
the negligible foreground contributions. 
$P$ was debiased as $\sqrt{P^2 - \Delta P^2}$ \citep{wardle74}, 
where $\Delta P$ was calculated from the measurement errors of photometry 
at four wave-plate angles. 
The polarization efficiencies are high, 95.5\%, 96.3\%, and 98.5\% at $J$, $H$, and $K$s, 
respectively \citep{kandori2006}, and no particular corrections were made here 
because of these high efficiencies. 
The measurable polarization of SIRPOL was reported to be $\sim$0.3\% all over the field of view 
at each band (Kandori et al. 2006), and so, $\Delta P=0.3\%$ was assigned to all the sources with  
$\Delta P <0.3\%$, which was derived simply from the photometric errors, at each band.
The 2MASS catalog \citep{skrutskie06} was used for photometric and astrometric calibration. 

\subsection{Source Selection for Analysis}

The sources detected at four wave-plate angles with photometric measurement 
errors of $<$0.1 mag  ($\Delta I_i/I_i \lesssim 0.1$, where $i$ denotes the wave-plate angle) were used for analysis, 
but the sources with $P^2 - \Delta P^2 \leqq 0$, i.e.,  non-polarization-detected sources, were not used.

\begin{figure}
 \begin{center}
 \FigureFile(60mm,*) {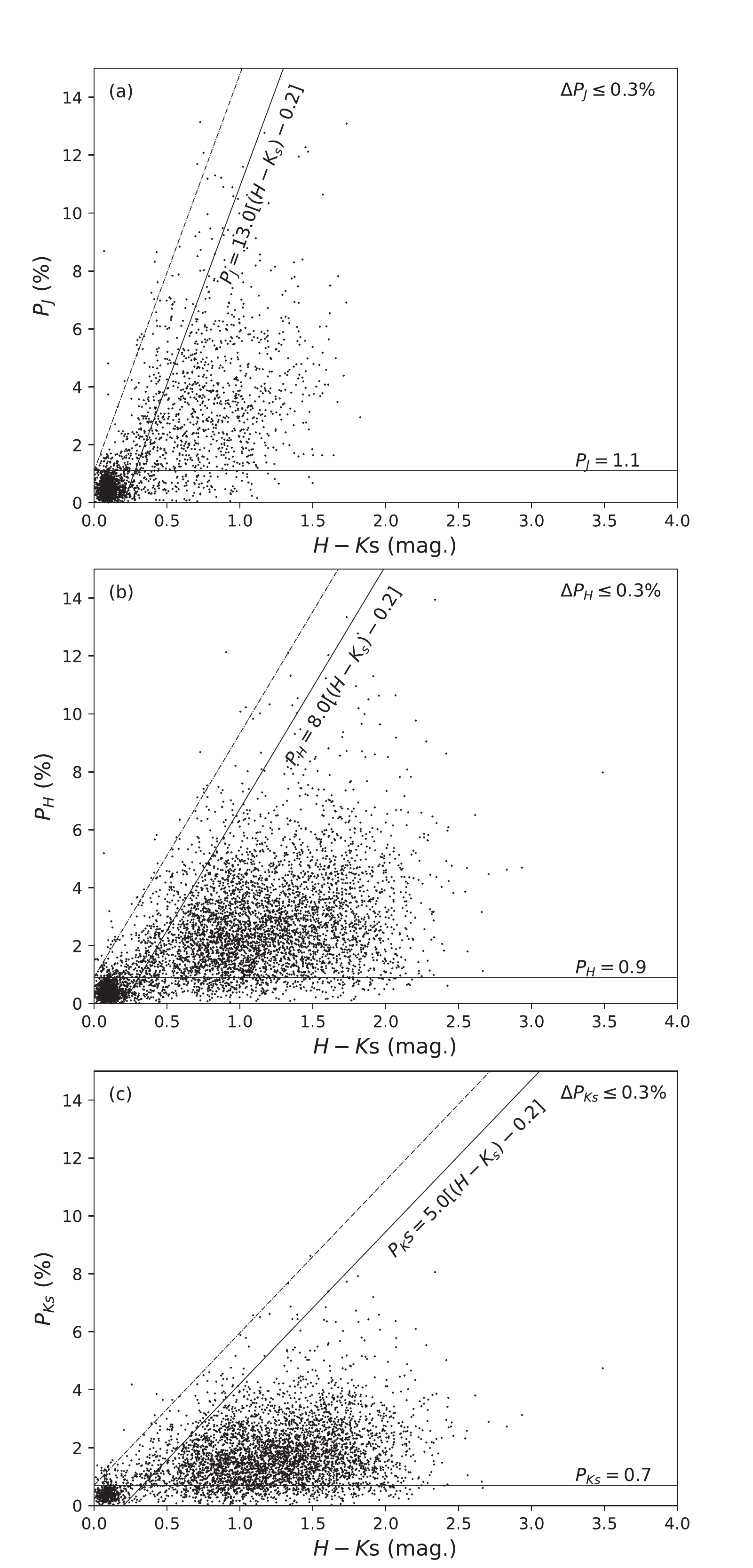} 
\end{center}
\caption{Polarization degree vs. $H - K$s color diagrams for the sources that are 
detected in all three bands ($J, H, and K$s) and have  $\Delta P$ of  $\leq0.3\%$ in each band.}\label{fig3}
\end{figure}

In order to exclude the sources with unusually large polarization degrees 
with respect to their $H - K$s color excess, we set an upper limit of the interstellar polarization, 
because the polarization of such source are not considered to be of dichroic origin.
To determine the upper limit at each band, only the good polarimetry sources 
with $\Delta P \leq 0.3\%$ are plotted in the polarization degree versus  
$H - K$s color diagrams (figure \ref{fig3}). 
We approximately estimate a threshold line to separate the outliers from the good measurements 
in each band (a dotted line at each panel).   
This line has a slope of 13 in J band, 8 in H band, and 5 in Ks band, respectively. 
Based on the model of  Galactic infrared point sources \citep{wainscoat1992} and 
on the sensitivity of SIRPOL, 
the expected colors of the background sources without the reddening 
by the M17 SWex cloud range mostly from $\sim0.2$ to $1.5$.
To ensure the sources for analysis to be located beyond M17 SWex, 
a solid line extending from the point of $H-K_{\rm s} =0.2$  and $P=0$
with the same slope as above is adopted as an upper limit line at each band. 

In addition, in order to exclude the comparatively low polarization sources 
with respect to their ($H-K$s) color excess, we set lower limits of the interstellar polarization.
In figure \ref{fig3}, high concentrations of plotted sources are seen near the lower left corners.
The sources of these concentrations can be identified in the DSS2 $R$-band image in almost all cases 
and are most likely to be foreground sources.  
In fact, almost all of their counterparts of the Gaia sources have the parallaxes 
that indicate their distances of $\lesssim2.0$ kpc \citep{gaia2018}. 
The background sources should have polarization degrees higher than those of the foreground ones.
Thus, we adopted lower limits of polarimetric degree of 1.1, 0.9, and 0.7 $\%$ in $J$, $H$, and $K$s bands, respectively. 
These limits are solid lines that run just above the upper edges of the concentrations.

\begin{figure}
 \begin{center}
  \FigureFile(60mm,*) {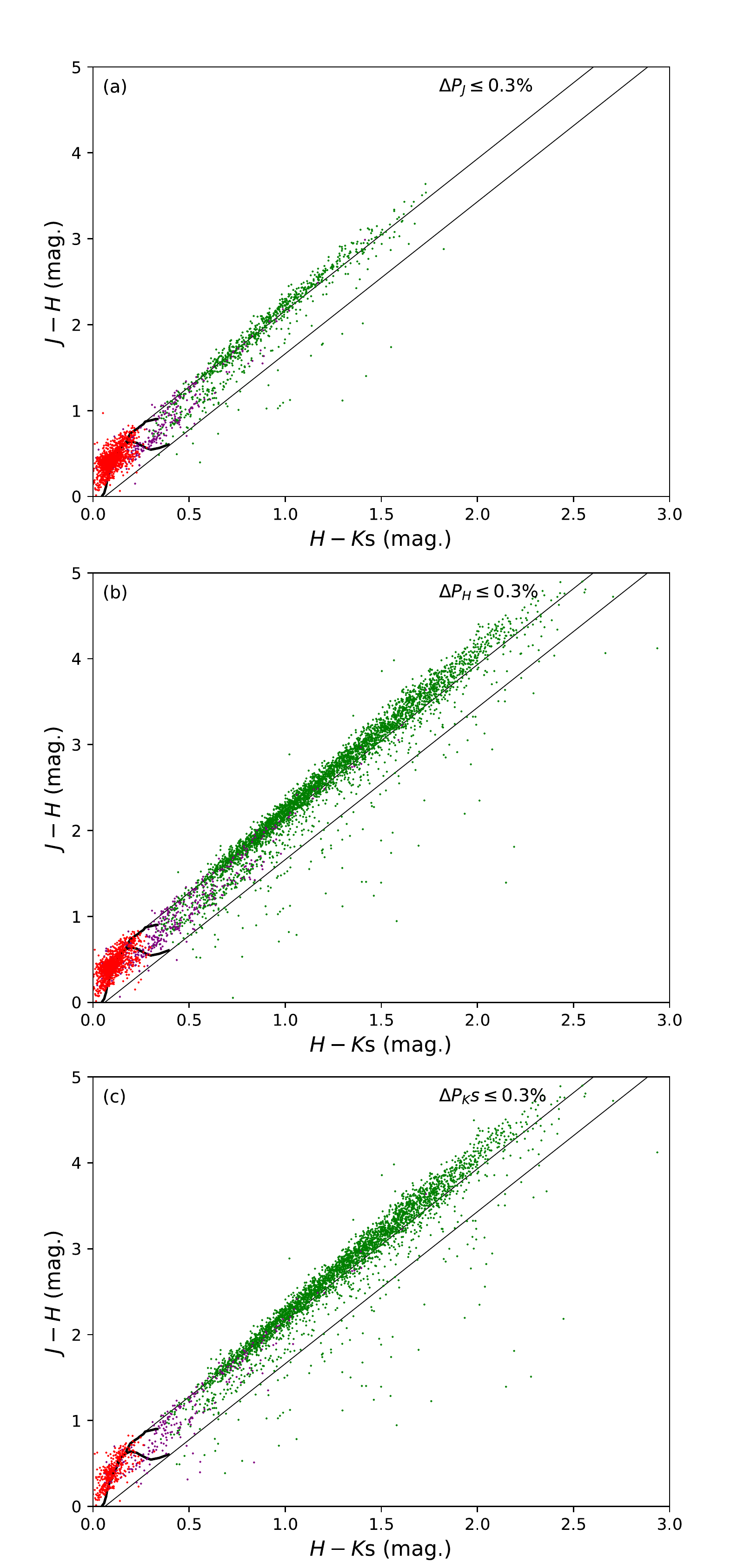} 
\end{center}
\caption{$J - H$ vs. $H - K$s color-color diagrams for the sources that are 
detected in all three bands ($J, H, K$s) and have $\Delta P$ of  $\leq0.3\%$ in each band. 
The solid, thick curves show the dwarf and giant loci \citep{bessell1988}. 
The thin lines are parallel to the reddening vector ($E(J -H)/E(H- K_{\rm s}) = 1.77$; \cite{nishiyama2009}). 
The colors for dwarfs and giants locus were transformed to those of 
the 2MASS system by using the equations of 2MASS website
${\rm (https://old.ipac.caltech.edu/2mass/releases/allsky/doc/sec6\_4b.html)}$. 
The notations of green, red, and purple marks are described in section 3.4.
}\label{fig4}
\end{figure}

The good polarimetry sources with $\Delta P \leq 0.3\%$ are also plotted 
in the $J - H$ versus  $H - K_{\rm s}$ color diagrams (figure \ref{fig4}.) 
The green marks are the sources located between the upper- and lower-limit lines 
in figure \ref{fig3} and most of them have the reddened colors of giants.
The red marks are the sources located near the lower left corner region fenced with the two limit lines 
in figure \ref{fig3} and are considered to be dwarfs with small extinction, i.e., nearby sources. 
The purple color marks are  the sources located above the two limit lines and below the dot-dash lines  
in figure \ref{fig3} and have the reddened colors of dwarfs. 
These with reddened dwarf colors possibly have some polarization information about  M17 SWex, 
but we didn't use them for analysis.

To ensure that the polarizations of the sources for analysis are caused by the M17 SWex cloud,  
further criterion was imposed.
We expect that the sources for analysis have ($H-K$s) color excess consistent with 
the H$_2$ column densities calculated in section 1.3.1, within an uncertainty of a factor of 2, 
by taking the the uncertainty of the column density estimate into account.  
($H - K_{\rm s})_0=0.2-1.5$  is adopted as the intrinsic color range of the background sources, 
because most of the background sources without the reddening by M17 SWex are 
expected to have an $H$-$K$s color range of $\sim0.2-1.5$, as mentioned above.
The column density from the ($H-K$s) color excess, $N_{E(H-K{\rm s})}$, is calculated 
with equations of $A_{\rm V}=E(H-K)/0.063$ (Rieke and Lebofsky 1985) 
and $N({\rm H}_2)/A_{\rm V}=1.0\times10^{21}$ cm$^{-2}$ (Lacy et al. 2017), 
i.e.,  $N_{E(H-K{\rm s})}=1.6\times10^{22}[(H-K{\rm s}) - (H-K{\rm s})_0]$ cm$^{-2}$. 
We excluded the sources that do not satisfy this consistency criterion. 

\section{Results}

\subsection{Near-IR polarization vector maps}

\begin{figure}
 \begin{center}
 \FigureFile(80mm,*) {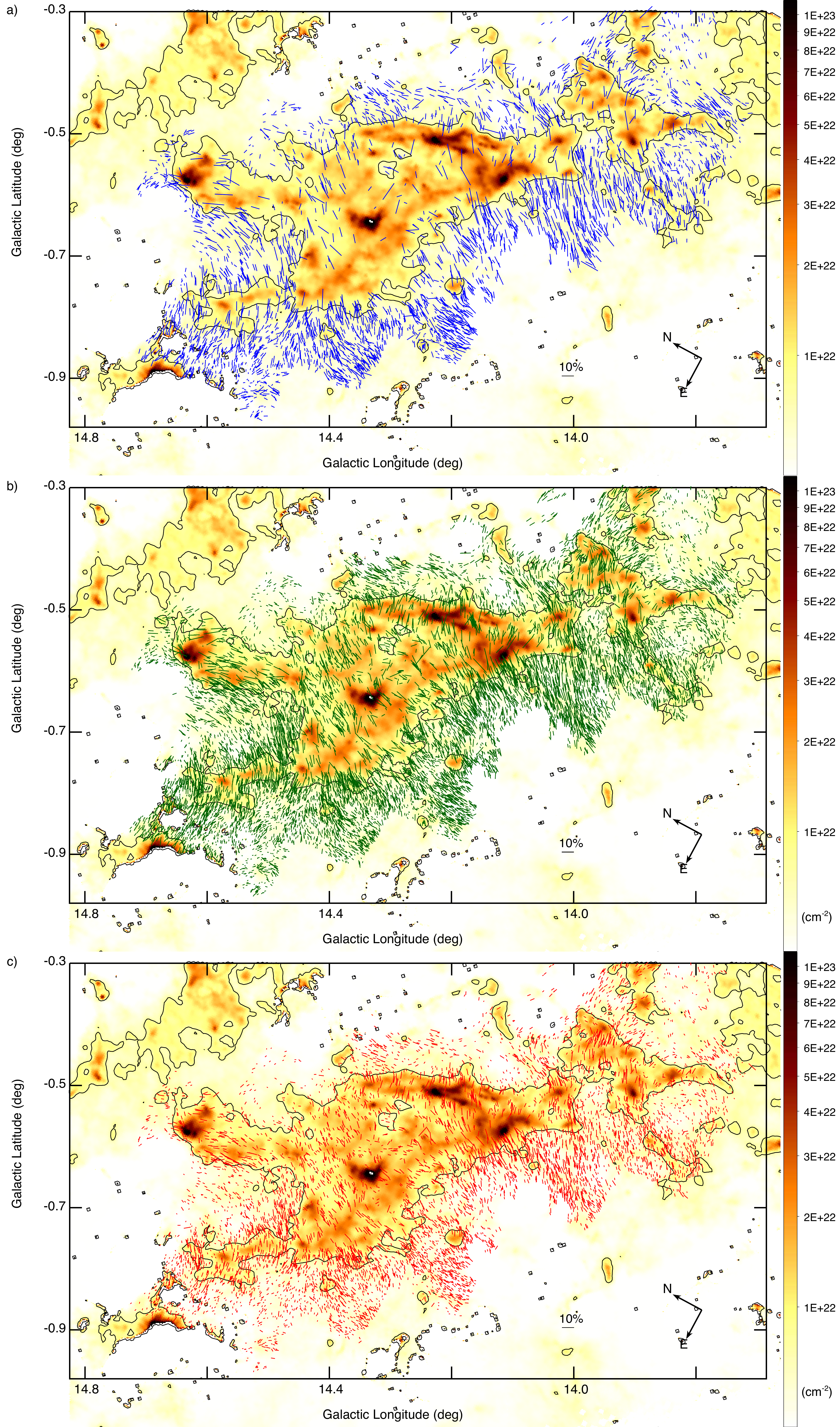} 
\end{center}
\caption{Near-IR polarization vector maps superposed on the H$_2$ column density 
map, which is the same as that in Figure 2. a) $J$-, b) $H$-, and c) $K$s-band vector maps, respectively.
The vectors are shown for the sources that mach the selection criteria described in section 3.4. 
The length of each vector is in proportion to it's polarization degree, and a vector of 10\% 
polarization is shown.
The contour lines of $7.0\times10^{21}$cm$^{-2}$, i.e., $A_{\rm V}=7$ mag., are shown by back, 
thin lines.}\label{fig5}
\end{figure}

\begin{figure}
 \begin{center}
 \FigureFile(60mm,*) {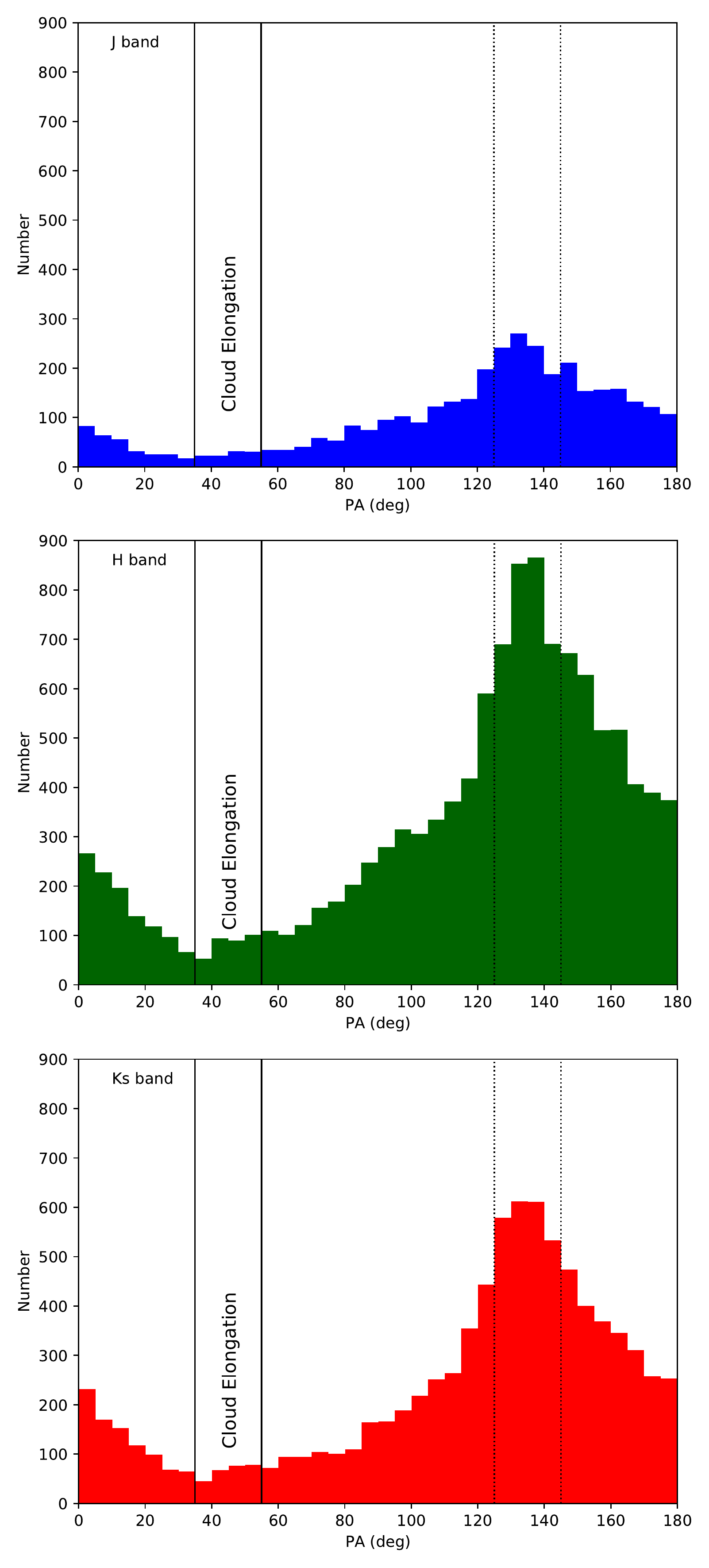} 
 \end{center}
\caption{Histograms of the polarization position angles of the sources that mach the selection criteria 
described in section 3.4.  An approximate position angle of  the cloud elongation  
for M17 SWex  is $\sim45\degree\pm10\degree$, which is shown by two solid lines, 
and its right angle by two dotted lines.}\label{fig6}
\end{figure}

The polarization vector maps and histograms of the polarization position angles 
for the $J$, $H$, and $K$s sources that have $P$ of $P/\Delta P > 3$ 
and match the selection criteria of section 3.4  
are presented in figures \ref{fig5} and \ref{fig6}, respectively.
In figure \ref{fig5} fewer vectors are seen at $J$ compared with $H$ and $K$s, particularly 
toward higher-column density regions with $N$(H$_2$)$>7\times10^{21}$ cm$^{-2}$. 
Filamentary substructures are seen within the higher-column density regions 
on the H$_2$ column density image (figure \ref{fig2}),  
which we treat in the next section. 
The distribution of the polarization vector angles has a peak of $\sim$130-135\degree, 
which is perpendicular to the M17 SWex cloud direction which is elongated in the NE-SW direction 
and its position angle is roughly estimated to be $\sim$35\degree-55\degree (figure \ref{fig6}).
This suggests a general trend that the global magnetic field is perpendicular to the cloud elongation as a whole.
However, the histogram spread widely around the peak, which also suggests some deviation 
from the general trend of the global magnetic field.

\begin{figure}
\begin{center}
 \FigureFile(80mm,*) {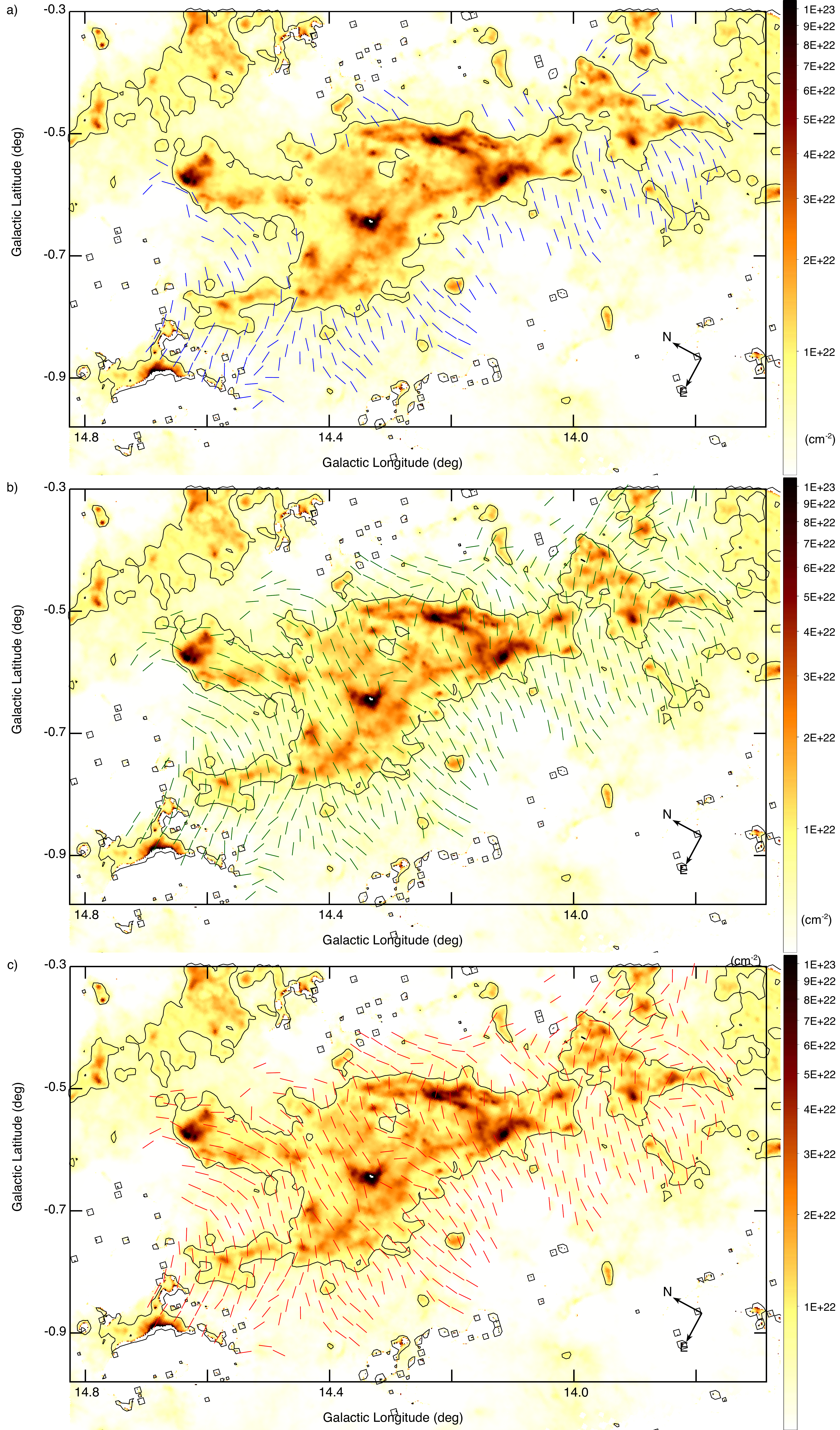} 
\end{center}
\caption{Averaged $J$, $H$, and $K$s polarization direction maps for the sources that mach the selection criteria 
of section 3.4, superposed on the H$_2$ column density map. 
The vectors only indicate the  averaged local polarization direction, but not polarization degree. 
}\label{fig7}
\end{figure}

To better understand the general trend and deviations,   
we averaged the position angles over an 1.\arcmin5-radius circle at grid points with an interval of 1.\arcmin5.    
The average of position angle $\overline{\theta}$ was calculated from the averages of 
$q'$ and $u'$ for the vector sources within each circle, 
i.e., $\overline{\theta} = (1/2)$tan$^{-1}(\overline{u'}/\overline{q'})$.
Here, only the averaged angle vectors in the areas where the number of vectors is 
greater than or equal to 10 are indicated in figure \ref{fig7}. 
These averaged vectors clearly show that the average magnetic field is ordered and smooth.  

\begin{figure}
 \begin{center}
  \FigureFile(80mm,*) {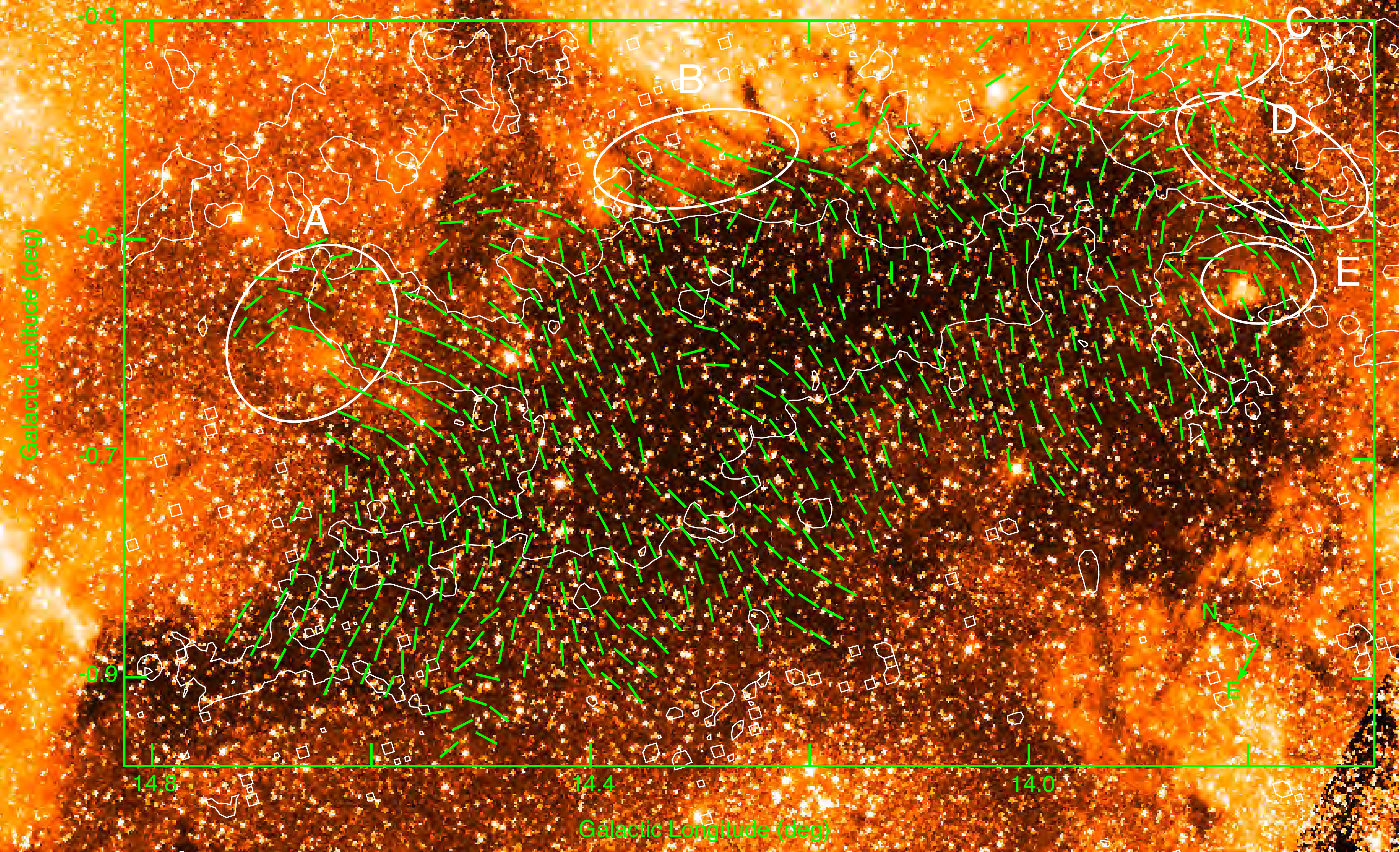} 
 \end{center}
\caption{Averaged $H$-band polarization direction map superposed on the DSS2 $R$-band image, 
which is emphasized by histogram equalization method. The vectors in the areas labeled with A—E 
may not indicate the magnetic field direction of the M17 SWex cloud.
}\label{fig8}
\end{figure}

To examine whether the field orientations sampled by our near-IR polarimetry are dominated 
by the magnetic fields in the volume of M17 SWex or by other portions of the line of sight,   
we overlaid the average $H$-band polarization vectors on the $R$-band images in figure \ref{fig8}.
Most of the vectors are located toward the dark region that can be recognized as the parts of the cloud, 
but some are located toward the optically bright areas, outside the dark region, 
i.e., toward the areas labeled with A—E in figure \ref{fig8}.
It is likely that the vectors located toward the bright areas (B—E) does not indicate the magnetic fields 
of the M17 SWex cloud, but the background fields.
There is a possibility that the area A is affected by the star-forming region M17, 
which is considered to be a nearby star formation region at the nearly same distance from the sun. 
The Planck 353 GHz polarization observations towards M17 SWex  \citep{planckXIX} show 
that the field orientation towards M17 SWex is different from that in its surroundings, 
where the field is parallel to the galactic plane.
The vectors in the areas of A-C might reflect the magnetic fields of the surroundings, 
which are parallel to the galactic plane, and contribute slightly to the spread in the histograms (figure \ref{fig6}). 
Excluding the average angle vectors toward the areas of A-E,  the remaining vectors indicate 
not only that the magnetic field is perpendicular to the cloud elongation as a whole, 
but also that at the both ends of the elongated cloud the magnetic field appears to bent toward its central part. 
This suggests a large-scale hour-glass shape of the magnetic field  
with a symetric axis angle of  PA$\sim130-135\degree$ in the M17 SWex cloud 
and the field bent at the both ends of the cloud is considered to contribute largely 
to the spread in the histograms, based on the local angular dispersions of the vectors, 
which are estimated to be not so large in the next section.  

\subsection{Comparison with C$^{18}$O ($J$=1-0) and $^{13}$CO ($J$=1-0) data}

The C$^{18}$O  and $^{13}$CO data \citep{nakamura2019a, shimoikura2019b, nguyen2019}
are used for comparison with the average polarization vector maps.

\begin{figure}
 \begin{center}
  \FigureFile(80mm,*) {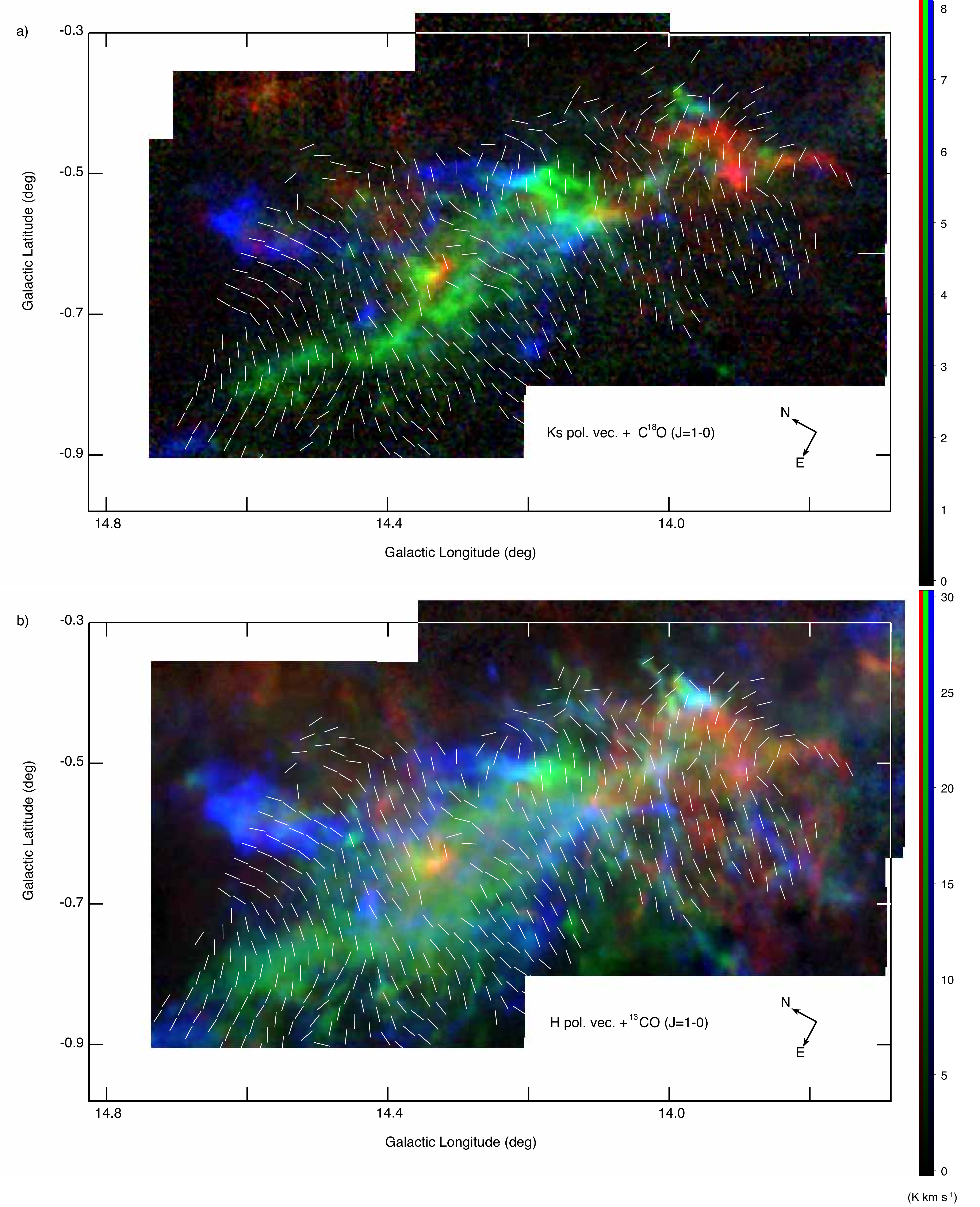} 
 \end{center}
\caption{Averaged polarization direction maps superposed on the C$^{18}$O ($J$ = 1—0) 
and $^{13}$CO ($J$ = 1—0) channel maps. 
Individual color scales are shown in units of K km s$^{-1}$  on the extreme rights of the figures.
a) Averaged $K$s polarization direction map superposed on the three-velocity channel images of C$^{18}$O 
( red: 22—25 km s$^{-1}$, green: 20—22km s$^{-1}$,  blue: 17—20 km s$^{-1}$).
b) Averaged $H$ polarization vector map superposed on the three-velocity channel images of $^{13}$CO, 
of which velocity ranges are the same as those of C$^{18}$O.
}\label{fig9}
\end{figure}

The $V_{\rm LSR}$ of $\sim$21 km s$^{-1}$ seems to be the central/system velocity of the M17 SWex cloud, 
judging from the C$^{18}$O data,  and we show three channel images of the following three velocity ranges 
(redshifted-, center-, and blueshifted-velocity ranges)  to roughly grasp its velocity structure.
In figure \ref{fig9}, three velocity channel images of C$^{18}$O and $^{13}$CO 
are shown in different colors: blue, green, and red represent the velocity-integrated intensity in the rage
17—20 km s$^{-1}$, 20—22 km s$^{-1}$, and 22—25 km s$^{-1}$, respectively.
As a general trend, the lower-longitude (SW) side of the cloud has redder velocities, 
and the higher-longitude (NE) side  has bluer velocities, although variabilities of the local velocities can be seen.  

In the C$^{18}$O images (figure \ref{fig9}a), the C$^{18}$O emission regions nearly correspond to 
the higher-column density regions of $N$(H$_2$)$\gtrsim7\times10^{21}$ cm$^{-2}$, 
and most of the individual features/filaments of C$^{18}$O seem to have their corresponding 
$N$(H$_2$) features/filaments in figure \ref{fig2}.  
Thus, the magnetic fields seem to be roughly perpendicular to most of the individual C$^{18}$O filaments as well.
On the other hand, in the $^{13}$CO images (figure \ref{fig9}b), such correspondences are not clear, 
but the possible envelopes of the individual C$^{18}$O/$N$(H$_2$) filaments seem noticeable  
toward the higher-column density regions.
In the outskirt (lower-column density) regions of $N$(H$_2$)$\lesssim7\times10^{21}$ cm$^{-2}$, 
many $^{13}$CO filamentary features protruded from the higher-column density regions can be recognized.  
These features seem to follow well the polarization angle vectors, i.e., magnetic fields.
These results are consistent with the well-known tendency that the main/dense filaments are 
perpendicular to the magnetic fields, while the sub/less-dense filaments are parallel (e.g., \cite{sugitani2011}). 
The Planck polarization study shows that the relative orientation 
between the column density gradient and magnetic field progressively changes 
with increasing column density from mostly parallel at $N({\rm H})>5\times10^{21}$ cm$^{-2}$ to 
mostly perpendicular at $N({\rm H})<5\times10^{21}$ cm$^{-2}$ \citep{planckXXXV}, 
i.e., in the higher-column density region the magnetic field perpendicular to the cloud elongation, 
while parallel in the lower-column density region. 
In the M17 SWex cloud, such magnetic field pattern exists but the orientation change seems to occur at the higher 
density of $N({\rm H}_2)\sim7\times10^{21}$ cm$^{-2}$, judging from the $^{13}$CO and C$^{18}$O data.

\section{Analysis and Discussions}

\subsection{Filamentary substructures in M17 SWex}

\begin{figure}
 \begin{center}
  \FigureFile(80mm,*) {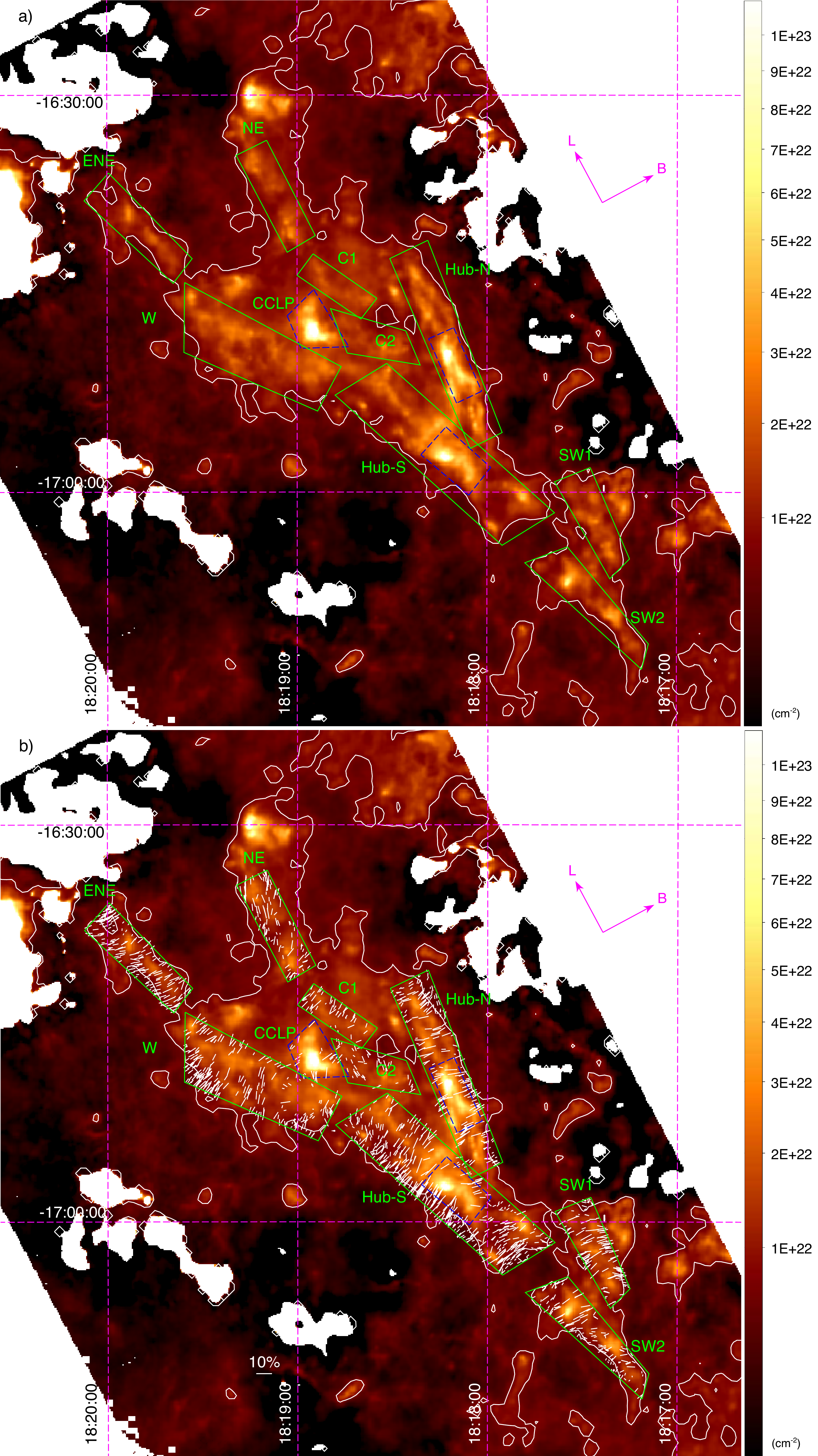} 
 \end{center}
\caption{Subregions on the H$_2$ column density map. North is at the top, east to the left. 
a) The subregions selected are shown by green quadrilaterals. 
Subregion CCLP and two sub-subregions in the Hub-N and Hub-S subregions are shown by blue dashed boxes. 
b) The $K$s-band polarization vectors for analysis are shown.}\label{fig10}
\end{figure}

We divided the M17 SWex cloud into several subregions to estimate the average column 
and volume number densities and the velocity dispersion in each subregion.
In the regions of $N$(H$_2$)$\gtrsim7\times10^{21}$ cm$^{-2}$ on the H$_2$ column density map, 
we subgrouped structures (filaments or clumps) that appear to be continuous or isolated by eye. 
We gave them their subregion names as SW1, SW2, Hub-N, Hub-S, C1, C2, W, NE, ENE, and CCLP, 
except a few clumps that are too small for analysis or seem to be affected from the external.
These subregions are shown by quadrilaterals in figure \ref{fig10}a.
Two sub-subregions that are the central parts of  the Hub-N and Hub-S subregions 
(Hub-N Center and Hub-S Center) are added to separately estimate the properties 
of these denser areas (dashed-line boxes in figure \ref{fig10}).  

The average column density $N({\rm H}_2)$ was obtained by averaging within each subregion, 
and the average number density $n({\rm H}_2)$ by simply dividing by a typical width of each subregion 
(table \ref{tab1}). 
The obtained volume number densities, $n({\rm H}_2)=(2.9\pm0.7)\times10^3$ cm$^{-3}$, 
except the CCLP subregion and two sub-subregions,  
are  smaller than the mean number density for the C$^{18}$O cores  
in the Orion A cloud \citep{ikeda2009} 
or the critical density of the C$^{18}$O line of $\sim5\times10^3$ cm$^{-3}$ 
(e.g., \cite{shimajiri2014}) and it is possible that the adopted width of each subregion is 
larger than the actual depth, i.e., the derived column density is a lower limit.
The velocity dispersion of C$^{18}$O  was obtained by single-gaussian fitting at each grid point 
of the C$^{18}$O  data cube, and these obtained dispersions within each (sub-)subregion are averaged. 
The standard deviations of the column densities and velocity dispersions of each (sub-)subregion are also 
shown in table \ref{tab1}.
The continuous structures of the subregions can be recognized to be elongated, except the CCLP subregion. 
We determined their directions of extension by fitting linear lines toward the data points of the subregions, 
where we adopted $1/N_{{\rm H}_2}$ as the data wight, and listed them with fitting error angles (table \ref{tab1}). 

\subsection{Magnetic fields and parameters for subregions}

The polarization vectors for the $K$s-band sources, 
which match the selection criteria in section 3.4 and are located within the subregions, 
are shown in figure \ref{fig10}b. 
Here, we use these $K$s-band vectors for analysis.

\begin{figure}
 \begin{center}
 \FigureFile(80mm,*) {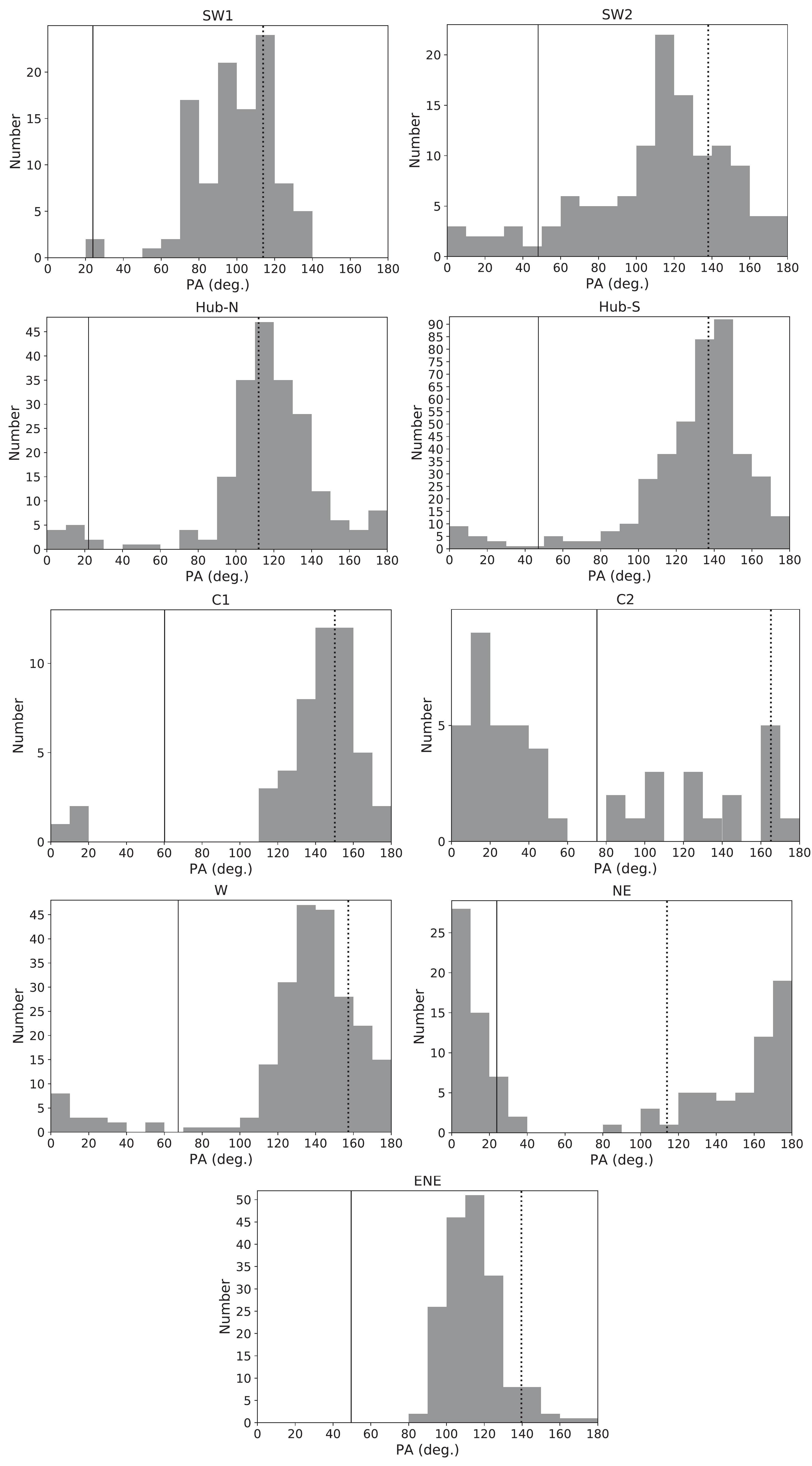} 
 \end{center}
\caption{Histograms of the polarization position angles of the sources within the subregions. 
 In each panel, an approximate position angle of  the elongation of each subregion is shown by a solid line, 
 and its perpendicular position angle by a dotted line.}\label{fig11}
\end{figure}

The histograms of the position angles for nine subregions, except the CCLP subregion, 
are presented in figure \ref{fig11}.
The best-fit elongation angle of each subregion is indicated by a solid line at each panel, 
and its perpendicular angle by a dotted line.
Most of the peak distribution angles of the subregions, except the C2 and NE subregions, 
are close to  those of these perpendicular lines and agree with them within $\sim$20-30$\degree$. 
This perpendicular trend is consistent with the fact that the cloud elongation is perpendicular 
to the magnetic field as a whole. 
The rather smaller scatters and slight deviation from the perpendicular angle of the ENE subregion 
might reflect the bent of the magnetic field toward the NE end region of the cloud.
The close angle to the elongation or parallel trend of the NE subregion also might reflect this bent.  

The strength of the plane-of-sky (POS) component of the magnetic field $B_{\rm pos}$ can be 
estimated, with the Davis-Chandrasekhar-Fermi method \citep{davis1951, chandrasekhar1953}, as 
\begin{equation}
  B_{\rm pos} = Q\frac{\sigma_{v}}{\sigma_\theta}\sqrt{4\pi \rho}, 
\end{equation}  
where $\rho$ is the mean volume density of the cloud, $\sigma_v$ is the velocity dispersion of line of sight, 
and $\sigma_\theta$ is the dispersion of the polarization vector angles, 
and $Q\sim0.5$ is a correction factor for $\sigma_\theta \lesssim 25\degree$ \citep{ostriker2001}.
We calculated the dispersion of the polarization angles with the Hildebrand method \citep{hildebrand2009}. 
The square of the differences in P.A. between the $N(l)$ pairs of vectors separated by displacements $l$, 
i.e., $\theta(l) \triplebond \theta ({\bf X}) - \theta({\bf X}+{\bf l})$, can be obtained by averaging them as 
\begin{equation}
  <\Delta\theta(l)^2>=\frac{1}{N(l)}\Sigma_{i=1}^{N(l)}[\theta ({\bf X}) - \theta({\bf X}+{\bf l})]^2,  
\end{equation}
where $< > $ denotes an average and $l=|{ \bf l}|$. 
In the case that the displacement $l$ exceeds the correlation length $\delta$,  
which characterizes a turbulent component $B_{\rm t}$, and is smaller than the scale $d$, 
which is the typical length scale for variations in the large-scale magnetic field  $B_0$, 
i.e., if $\delta < l \ll d$, the square of the total measured dispersion function is expected to become 
\begin{equation}
  <\Delta\theta(l)^2> _{\rm tot} \simeq b^2 + m^2 l^2+\sigma_{\rm M}^2(l),  
\end{equation}
where $m$ is a coefficient that indicates the contribution from $B_0$, $\sigma_{\rm M}(l)$ 
is a measurement uncertainty,  and $b^2$ is the intercept of the squared function at $l=0$.
Here, $b$ is linked to $B_{\rm t}$ as 
\begin{equation}
  \frac{<B_{\rm t}^2>^{1/2}}{B_0} =\frac{b}{\sqrt{2-b^2}}, 
\end{equation}
and we derived the dispersion angle $\sigma_\theta$ as $\sigma_\theta=< B_{\rm t}^2>^{1/2}/B_0$
in the same manner as \citet{chapman2011}.

\begin{figure}
 \begin{center}
  \FigureFile(80mm,*) {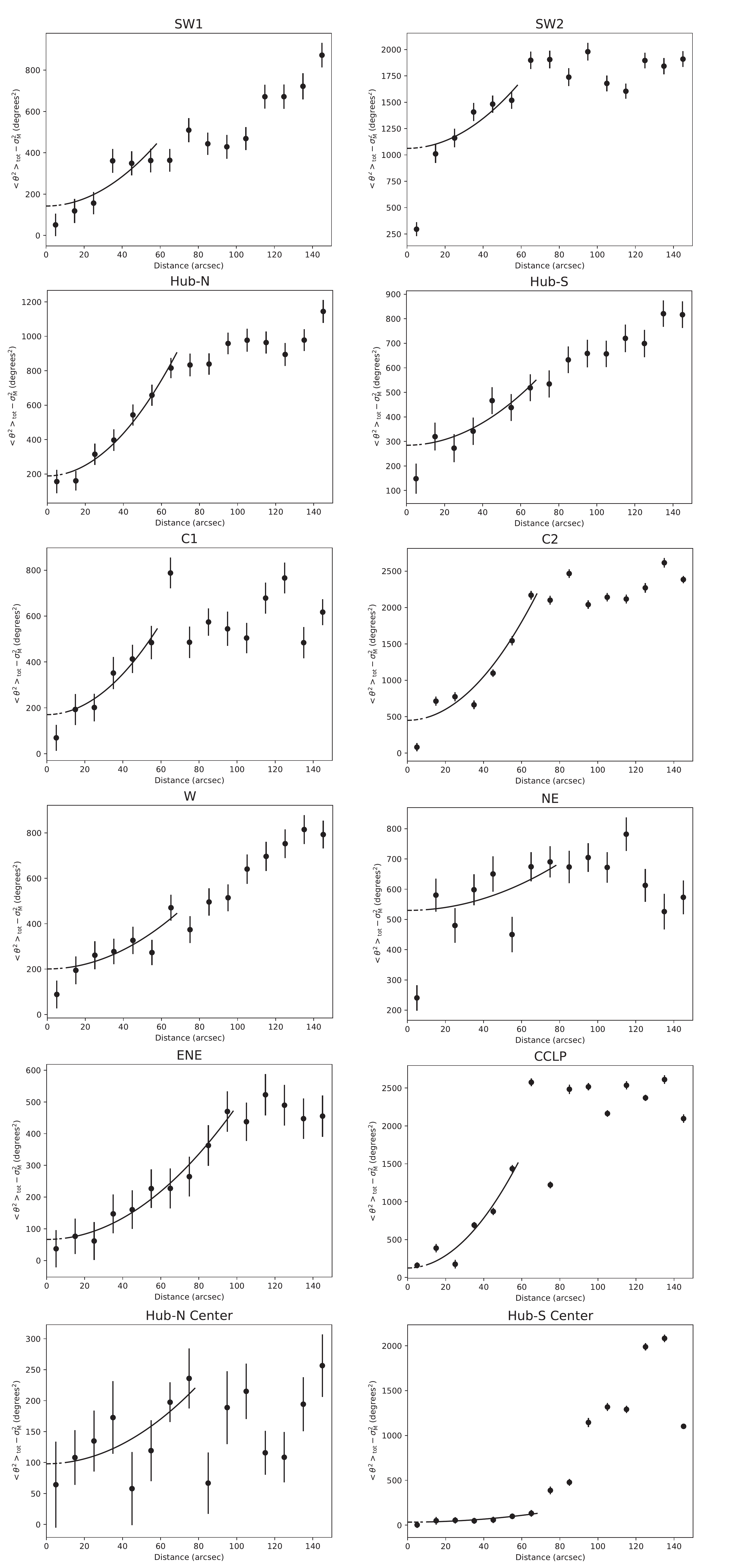} 
 \end{center}
\caption{Square of the total  measured dispersion function versus displacement distance for the (sub-)subregions. 
The square of the measurement  uncertainty $\sigma_{M}^2$ is subtracted from the square of 
the the total  measured dispersion function. 
The best-fit power law line is shown by a solid line in each panel and $b^2$ is estimated at the intercept point of the displacement distance of $l=0$.}\label{fig12}
\end{figure}

We fitted the polarization data with the squared function at each (sub-)subregion  
with a bin width of $10\arcsec$ ($\sim0.1$ pc at $d=2.0$ kpc) without use of 
the data of the first bin ($0<l<10\arcsec$) in a similar way to \citet{chapman2011} 
and showed the fitting results in figure \ref{fig12}. 
The $b$ of each (sub-)subregion and the individual vector number for fitting are presented in table 2, 
and all of the $\sigma_\theta$ derived from the $b$'s are $<25.0\degree$.
The derived strengths of plane-of-sky component of the magnetic field $B_{\rm pos}$ range 
from $\sim$70 to 300~$\mu$G in the subregions.
The strengths of the C1 and CCLP subregions are $\sim$2—3 time larger than those of $\sim$100~$\mu$G 
in the other subregions. 
The larger strength of the C1 subregion might be due to its larger velocity dispersion 
with respect to it's moderate dispersion angle. 
The CCLP subregion seems to have a higher number density, like the two sub-subregions,  
and this might caused the larger strength as a result.
However, the uncertainty of $b$ is much larger than the others 
probably because of its small number of vectors. 
Note that \citet{sato2010} assigned a distance of 1.1$\pm$0.1 kpc for the H$_2$O maser source 
situated toward the G14.33-0.64 cloud, which may correspond to the CCLP subregion, 
suggesting that it is a foreground cloud. 
If we adopt 1.1 kpc as a distance of the CCLP subregion, $B_{\rm pos}$ becomes  $\sim$210 $\mu$G. 
However, the peak C$^{18}$O velocity of $\sim$22 km s$^{-1}$ at the G14.33-0.64 position might suggest 
that the CCLP subregion belongs to the M17 SWex cloud.
\citet{santos2016} also estimated the magnetic field strengths toward our defined sub-subregions 
of Hub-N Center and Hub-S Center and reported the POS strengths  of 250 $\mu$G and 430 $\mu$G, respectively. 
Although the estimating method and areas are slightly different from ours, their derived values are rather consistent 
with our ones. 

We evaluated the dynamical states of the magnetized (sub-)subregions,  
with the ratio of the cloud mass to the magnetic flux, i.e., the mass-to-flux ratio. 
The critical mass-to-flux ratio for the magnetic stability, $\lambda_{\rm crit}=1/\sqrt{4\pi^2G}$, is given  
by \citet{nakano1978}, where $G$ is the gravitational  constant. 
We calculated the dimensionless mass-to-flux ratios 
normalized by  the critical mass-to-flux ratio 
as follows; 
\begin{equation}
 \lambda_{\rm obs}=\frac{\lambda_{\rm pos}}{\lambda_{\rm crit}}=\frac{\mu m_{\rm H}}{\lambda_{\rm crit}}\frac{N({\rm H}_2)}{B_{\rm pos}}, 
\end{equation}
where $\mu$ is the mean molecular weight of 2.8, and $m_{\rm H}$ is the hydrogen mass.
The derived normalized ratios $\lambda_{\rm obs}$ are shown in the last column of table 2 and their average is $1.1(\pm0.4)$, 
where the number in parentheses indicates their standard deviation 
(note that $\lambda_{\rm obs} \sim1.3$ for  CCLM at $d=1.1$ kpc).
Here only the magnetic fields of plane-of-sky are taken into account, but those of line-of-sight should be in existence. 
In practical ways the magnetic field of line-of-sight should also be taken into account. 
For example, if we simply assume a moderate inclination angle of the magnetic field of 45\degree to the plane of sky,  
the total normalized ratio would decrease to $\sim0.7$ times.  
\citet{crutcher2004} mentioned that a statistical mean correction for geometrical bias 
may reduce the observed ratio by down to 1/3.
Thus, the total normalized ratio $(\lambda_{\rm pos + los}/\lambda_{\rm crit})$  are expected to be 
smaller than the observed normalized ratio $ \lambda_{\rm obs}$. 
These suggest that the (sub-)subregions are in the magnetically critical or sub-critical states.  
For the Hub-N Center and Hub-S Center sub-subregions, the tendency of the magnetic state seems consistent 
with that of \citet{santos2016}.

\subsection{Role of multi-scale magnetic fields in star-forming regions}

Our results strongly suggest that the shape and evolution of the M17 SWex cloud is controlled 
by the magnetic fields. 
This would be also supported by the smooth field structure, as described in \citet{heiles2005}.
As described in section 4, the polarization map implies that the magnetic field is globally perpendicular 
to the cloud elongation of the higher-column density parts, 
and that the local magnetic fields are also perpendicular to most of the individual filamentary subregions  
within the higher-column density parts with the magnetically stable conditions. 
Moreover,  the $^{13}$CO filamentary features protruded 
from the higher-column density parts seem to follow the magnetic fields (Figure 10) 
and may be candidates of feeding filaments onto the higher-column density parts. 

\citet{soler2013} first introduced Histograms of Relative Orientation (HRO), 
which shows the correlation between the density gradient of the cloud structure and the magnetic field direction, 
and applied for their simulated data of molecular clouds.  
They suggested that in magnetized clouds density gradients preferentially become 
from perpendicular to parallel to magnetic field in regions with density over some critical/threshold density 
and this change is also presented in the projected maps, i.e., column density maps.  
Their further study \citep{soler2017b} suggested that the observed change in the HRO analysis  is 
as the result of the gravitational collapse and/or convergence of flows 
and the density threshold is  related to the magnetic strength. 
In a paper that present a new technique to trace magnetic fields in turbulent media 
using spectral line data \citep{lazarian2018}, they mentioned that in the self-gravitating regions 
both of the density and velocity gradients are expected to change their alignment 
parallel to the magnetic field and the change of the density gradient happens 
earlier than that of the velocity gradient. 
Based on these works, it is most likely that in M17 SWex self-gravity is relatively dominant 
in the higher-column density parts, 
while not in the lower-column density parts, i.e., the $^{13}$CO filament area.

The magnetic field lines try to support the cloud to prevent the collapse 
and form the hourglass shape morphology as a result. 
Under such magnetically stable condition of the cloud as a whole, 
the quick collapse that leads to high-mass star formation might not be expected, 
even so the cloud has a large mass enough for high-mass star formation. 
Only small, local areas within the filaments might have been able to form intermediate- and lower-mass stars 
with the aid of some magnetic diffusion, such as an ambipolar diffusion and turbulence reconnection. 
Star formation by the magnetic diffusion might take more time than that occurred in the unstable condition 
(e.g., \cite{nakamura2008}) and continue for a longer time, 
i.e., the wide spatial and temporal distributions of stars could be expected. 
Thus, the magnetically stable conditions appear consistent with the present low-level of 
high-mass star formation and the continuous ($>$ 1 Myr) formation of intermediate-mass stars reported 
by \citet{povich2010}. 
 
As described in section 4.2, the lower-longitude (SW) side of the cloud has redder velocities and 
the higher-longitude (NE) side has bluer velocities. 
This velocity structure might be interpreted as the global velocity gradient along the cloud elongation 
and suggests the large-scale contraction along the cloud elongation. 
The velocity gradients along some IRDC filaments were reported \citep{tackenberg2014,nguyen2011} 
and interpreted them as gas flows along the filaments, i.e., accretion flows onto the densest regions.
If the global velocity gradient in the M17 SWex cloud is really due to global gas flow along the cloud elongation, 
the bent magnetic fields on the both ends of the cloud or large-scale hourglass-shape magnetic field  
would be the result of the contraction along the elongation toward the central part of the cloud. 
This type of the large-scale bent can be also seen in other molecular clouds. 
For example, in case of the Serpens South cloud, it is reported that the magnetic field 
on the southern side of the Serpens South protocluster is clearly curved toward the cluster 
(figures 6-8 of \cite{sugitani2011}), suggesting the idea that the magnetic field is distorted 
by gravitational contraction along the main filament toward the cluster. 
The velocity gradient along the main filament (\cite{kirk2013, tanaka2013, nakamura2014, shimoikura2019a}) 
might support this type of the large-scale field distortion (see also figure 9a of \cite{andre2014}).
In another instance, in the Orion A molecular cloud, which appears similar to this cloud in morphology, 
the magnetic field is nearly perpendicular to the cloud elongation toward the northern and mid parts, 
while the curved-shape field can be seen toward the molecular gas of its southern end
(figure 5 of \citep{planckXXXV}). 

\begin{figure}[h]
  \begin{center}
  \FigureFile(80mm,*) {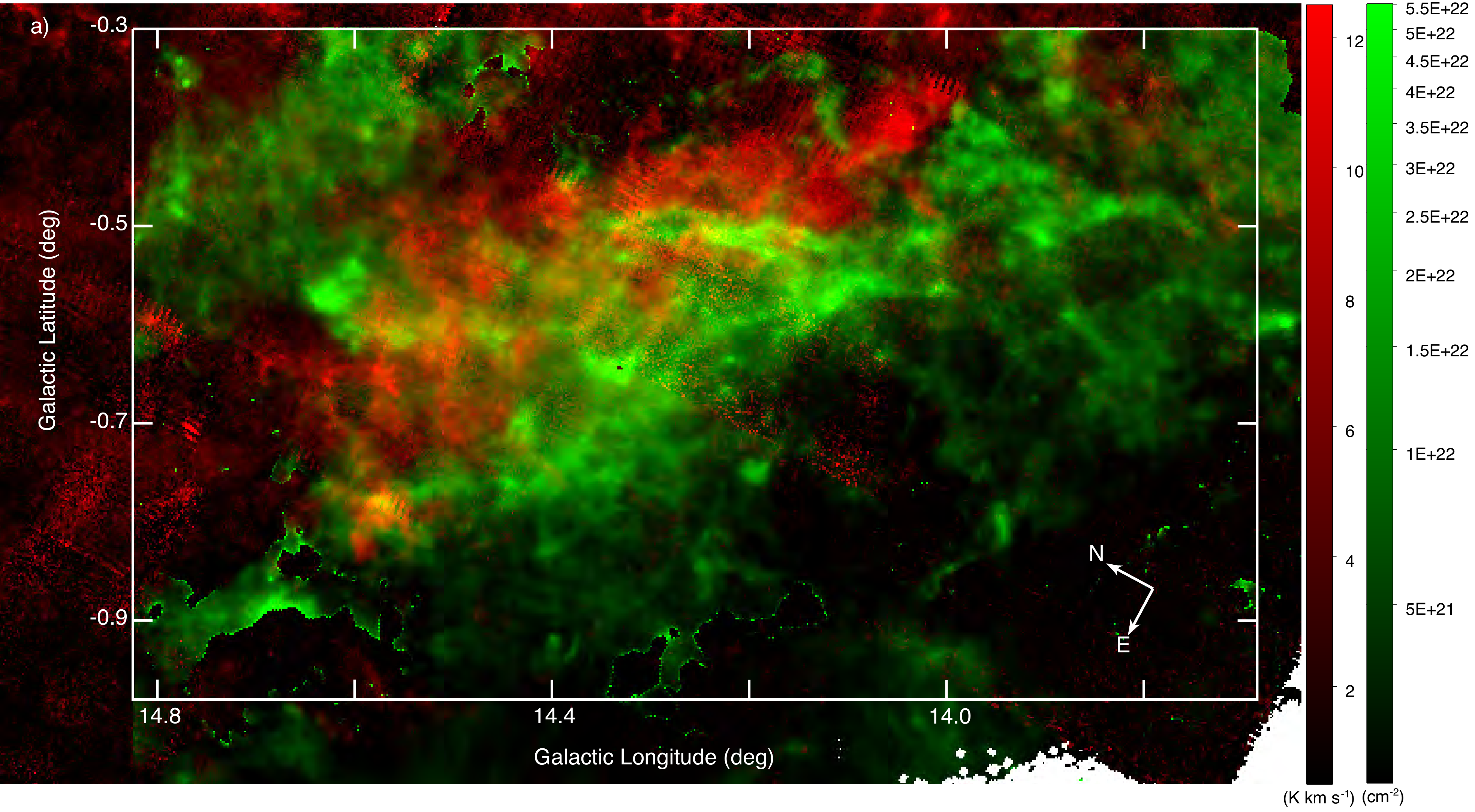} 
  \end{center}
 \caption{CO ($J$ = 3—2)  intensity image integrated over the velocity range
 $V_{\rm LSR}$ = 32.8—36.8 km s$^{-1}$ (red) and the H$_2$ column density image (green), 
 which was derived in section 3.1 by SED fitting of the Herschel archival data (160, 250, 350, and 500 $\mu$m). 
 This CO feature with a peak velocity of $\sim35$ km s$^{-1}$ is $\sim14$ km s$^{-1}$ redshifted 
 from the central velocity of the M17 SWex cloud of $\sim21$ km s$^{-1}$,  
 and the distributions of this feature and the M17 SWex cloud are in good agreement.  
 This positional agreement may not be by chance along the line of sight. }\label{fig13}
\end{figure}

The 3D structure (or the depth) of the M17 SWex cloud is unknown, 
but we guess, from less overlapping appearances of the individual filamentary subregions 
within the higher-column density parts, that the M17 SWex cloud has a sheet-like cloud structure as a whole 
and that we look at this structure with a moderate angle of inclination.
If the sheet-like structure is real, two magnetic field configurations can be possible. 
One is perpendicular to the sheet-like structure, the other is parallel to it. 
The parallel configuration can be expected when the sheet-like structure was created by external large-scale flows  
such as an expansion of H II region (e.g., \cite{shimoikura2015}, 2019), 
supernovae (e.g., \cite{tatematsu1990}; \cite{matsumoto2015}; \cite{dobashi2019a}), 
or superbubbles (e.g., \cite{inutsuka2015}). 
Since the M17 SWex cloud is close to the Galactic plane, such flows can be generally expected.   
In effect, the magnetically stable states of the subregions could be caused 
by the enhancement  of the magnetic field strength due to the compression of some external flow.
Therefore, we searched the CO ($J=3-2$) archival data of JCMT for flow candidates and found one possible candidate.
Figure \ref{fig13} is an integrated intensity map between 32.8 km s$^{-1}$ and 36.8 km s$^{-1}$ in $V_{\rm LSR}$, 
which indicates a CO component of $V_{\rm LSR} \sim35$ km s$^{-1}$ and is much weaker than 
the main component  of the M17 SWex cloud of $\sim21$ km s$^{-1}$. 
The distribution of this component and the higher-latitude side of the higher-column density parts 
of the M17 SWex cloud are in very good agreement, which would not be by chance just along the line of sight.  
There could be some relationship between these two, e.g., 
a remnant of a large-scale flow that had passed through the cloud  
or a shock compressed layer due to a large-scale flow, and etc. 
This $\sim35$ km s$^{-1}$ component will be discussed in more detail 
by \citet{nguyen2019} and \citet{shimoikura2019b}.   

Finally, we speculatively discuss the perpendicular configuration of the magnetic field. 
While the global magnetic field in the Galactic plane is roughly parallel to the Galactic plane, 
the magnetic fields of M17 SWex look perpendicular or oblique to the Galactic plane. 
As mentioned before, this can be also confirmed by the Planck 353-GHz 
polarization observations \citep{planckXIX}. 
However, it may be difficult to achieve such magnetic configuration from the global gravitational instability, 
which is unstable only to the symmetric mode relative to the plane \citep{hanawa1992}. 
Thus, the global field should be roughly parallel to the plane if the cloud is formed 
by the global gravitational instability. 
The oblique or perpendicular configuration can be achieved by the Parker instability \citep{parker1966}  
that is most unstable to the anti-symmetric mode \citep{horiuchi1988}, 
which can make the field lines cross the Galactic plane. 
If this is the case, the M17 SWex cloud may be formed at the foot part of the large-scale magnetic field 
floated up from the Galactic plane by the Parker instability.
If gas flows came from a place away from the Galactic plane along the magnetic field toward the Galactic plane, 
the sheet-like structure perpendicular to the magnetic field could be produced by the shock compression 
due to the rapid deceleration of gas flows near the Galactic plane. 
The M17 SWex cloud has a sharp column-density gradient on the near side of the Galactic plane, 
while a gentle gradient on the far side, of which the appearance could be expected by the Parker instability.
Moreover, a CO component of $V_{\rm LSR} \sim 35$ km s$^{-1}$ could be interpreted as a flow that comes later 
along the floated-up magnetic field and collides with the main cloud component with $\sim$21 km~s$^{-1}$. 
The violent event such as the cloud-cloud collision along the magnetic field seems to be consistent 
with the dynamical states of the clumps in M17 SWex.  
The Planck 353-GHz polarization observations show that the oblique or perpendicular configuration of the magnetic field last down to $b\sim-1.3$\degree ~in this region beyond our observation coverage and 
the similar configuration is seen toward the Orion region. 
Further studies are needed concerning this kind of the magnetic field configurations, 
oblique or perpendicular to the galactic plane, in the context of molecular cloud formation/evolution.  
 

\section{Summary}

We performed $JHK$s imaging polarimetry of the M17 SWex cloud  
and revealed the overall magnetic field structure and the local ones of 
individual filamentary subregions that are situated in the regions of higher-column density. 
The Herschel archival data and the NRO 45~m telescope data were also used for our analysis. 
The main findings are as follows: 

\begin{enumerate}
  \item The global magnetic field of the cloud is perpendicular to its elongation as a whole, 
   and the magnetic fields at the both ends of the elongated cloud appear to bent toward its central part. 
   This suggests a large-scale hour-glass shape of the magnetic field in the M17 SWex cloud. 
   The large-scale bent is likely due to the molecular gas contraction along the cloud elongation.
   
  \item The major filamentary structures were picked up as our defined subregions 
  in the higher-column density regions of $N$(H$_2$)$\gtrsim7\times10^{21}$ cm$^{-2}$ 
  on the H$_2$ column density map, which was produced by the Hershel data.
  The local magnetic fields in the filamentary subregions are mostly perpendicular to their elongations. 
  At the NE end of the cloud, the local fields of the ENE and NE subregions are 
  slightly oblique or nearly parallel, respectively, reflecting the large-scale bent of the magnetic field. 
  
  \item The  C$^{18}$O ($J$=1-0)  and $^{13}$CO ($J$=1-0) data show the velocity gradient along the cloud elongation, 
  which could suggest the gas contraction along the cloud elongation. 
  The C$^{18}$O  filamentary structures seem identical to the subregions defined on the H$_2$ column density map. 
  The $^{13}$CO  filamentary structures, which protrude from the higher density regions, 
  seem to follow the magnetic field in the outskirt of the cloud.
  
  \item The magnetic field strength was derived with the Davis-Chandrasekhar-Fermi method in each subregion. 
  The derived strengths of the plane-of-sky component of the magnetic field range from $\sim$70 to 300 $\mu$G,  
  typically $\sim$100 $\mu$G, in the subregions. 
  The magnetic stability of the magnetized subregions was evaluated and it is found that the subregions are likely 
  to be magnetically critical or sub-critical. 
  These magnetically stable subregions appear consistent with the present low-level of high-mass star formation 
  and the longterm formation of intermediate-mass stars, which were reported in the M17 SWex cloud. 
\end{enumerate} 

\begin{ack}
We thank the anonymous referee for reading the paper carefully and providing useful comments.
This work was partly supported by JSPS KAKENHI Grant Numbers JP24540233, JP16H05730 and JP17H01118. 
K.S. thanks Y. Nakajima for assistance in the data reduction with the SIRPOL pipeline package. 
The IRSF project is a collaboration between Nagoya University and the South African Astronomical Observatory (SAAO) 
supported by the Grants-in-Aid for Scientific Research on Priority Areas (A) (No. 10147207 and No. 10147214) 
and the Optical \& Near-Infrared Astronomy Inter-University Cooperation Program from MEXT Japan 
and the National Research Foundation (NRF) of South Africa. 
M.T. is partly supported by JSPS KAKENHI Grant Numbers 18H05442 and 15H02063.
\end{ack}

\draft

\begin{table}
  \tbl{Approximate Sizes and Averaged Physical Parameters for Subregions.}{%
  \begin{tabular}{lccccc}
      \hline
      Subregion/Sub-subregion & Approximate Size & $N$ (H$_2$) & $n$ (H$_2$) & $\sigma_v$ &  Direction of extentson\\ 
          & (pc$\times$pc) &  ($\times 10^{22}$cm$^{-2}$) & ($\times10^{3}$cm$^{-3}$) &  (km s$^{-1}$) & (deg. in PA)\\
      \hline
      SW1    & 4.6$\times$1.4 &1.29$\pm$0.46  & 2.9  & 0.85$\pm$0.26 & 23.9$\pm$0.4\\
      SW2    & 6.1$\times$1.5 &1.34$\pm$0.56  & 2.9 & 1.45$\pm$0.59 & 48.1$\pm$ 0.3\\
      Hub-N  & 9.1$\times$1.7 &1.97$\pm$1.01 & 3.8  & 0.68$\pm$0.17& 21.8$\pm$0.2\\
      Hub-S  & 9.7$\times$2.6 &1.63$\pm$0.82 & 2.0  & 1.22$\pm$0.33 & 45.1$\pm$0.2\\
      C1        & 3.3$\times$1.2 &1.45$\pm$0.29 & 4.1  &  1.76$\pm$0.54 & 60.2$\pm$0.6\\
      C2        & 3.4$\times$1.5 &1.31$\pm$0.39 & 2.8  &  1.17$\pm$0.34 & 75.2$\pm$1.4\\
      W         & 7.2$\times$2.4 &1.43$\pm$0.54 & 1.9  &  0.93$\pm$0.24 & 67.3$\pm$0.3\\
      NE       & 4.8$\times$1.4 &1.35$\pm$0.44 &  3.1 &   0.91$\pm$0.22 &23.9$\pm$0.4\\
      ENE    &  5.3$\times$1.5 &1.00$\pm$0.32 &  2.2 &  0.68$\pm$0.17 & 49.6$\pm$0.4\\
      CCLP (Center CLumP) & 2.2$\times$1.7 & 3.58$\pm$1.34 & 6.9 & 1.23$\pm$0.24 & ...\\
       \hline
      Hub-N Center & 3.1$\times$1.2 & 3.58$\pm$1.50 & 9.5 & 0.74$\pm$0.12 & ...\\
      Hub-S Center & 2.7$\times$1.5 & 3.00$\pm$1.24 & 6.6 & 1.21$\pm$0.28 & ...\\
      \hline
    \end{tabular}}\label{tab1}
\end{table}

\begin{table}
  \tbl{Magnetic Parameters for Subregions.}{
  \begin{tabular}{lcccc}
      \hline
      Subregion/Sub-subregion & $b$ & Number of &  $B_{\rm POS}$& $\lambda_{\rm obs}$ \\ 
      		      &  (deg) & Vectors & ($\mu$G)  & \\
      \hline
      SW1   &  13.3$\pm$2.2 & 104 & 110$\pm$40 & 0.92$\pm$0.36\\
      SW2   &  32.3$\pm$1.0 & 123 &   70$\pm$32 & 1.45$\pm$0.66\\ 
      Hub-N &  13.7$\pm$1.1 & 209 &   94$\pm$34 & 1.59$\pm$0.58\\
      Hub-S &  16.9$\pm$0.9 & 420 &   99$\pm$37 &  1.25$\pm$0.46\\
      C1       &  13.0$\pm$1.1 &   49 &  260$\pm$90 &  0.42$\pm$0.14\\
      C2       &   21.2$\pm$3.0 &  47 &   87$\pm$32 &  1.14$\pm$0.41\\
      W        &  14.2$\pm$1.3 & 227 &  88$\pm$29 &  1.24$\pm$0.42\\
      NE      &   23.0$\pm$1.2 & 107  &  66$\pm$20 &  1.56$\pm$0.47\\
      ENE   &     8.1$\pm$0.9 & 178  & 120$\pm$40 & 0.62$\pm$0.20\\
      CCLP (Center CLumP) &  11.2$\pm$5.9 & 26 & 280$\pm$170 & 0.97$\pm$0.58\\
       \hline
      Hub-N Center & 9.9$\pm$1.6 & 30 & 220$\pm$70 & 1.21$\pm$0.37\\
      Hub-S Center &  5.8$\pm$0.8 & 31 & 530$\pm$180 & 0.43$\pm$0.15\\
          \hline
    \end{tabular}}\label{tab2}
\end{table}

\newpage


\begin{thebibliography}{}
\bibitem[Andersson et al.(2015)]{andersson2015}
 Andersson, B. G., Lazarian, A., \& Vaillancourt, J. E., 2015, \araa, 53, 501
\bibitem[Andre et al.(2014)]{andre2014}
 Andre, P., Di Francesco, J., Ward-Thompson, D., Inutsuka, S., Pudritz, R. E., \& Pineda, J. E., 2014, 
 In Protostars and Planets VI, ed. H. Beuther, et al., 
 University of Arizona Press, Tucson, pp. 27-51. 
\bibitem[Aniano et al.(2011)]{aniano2011}
 Aniano, G., Draine, B. T., Gordon, K. D., \&  Sandstrom, K. 2011, \pasp, 123, 1218 
\bibitem[Arce et al. 2010]{arce2010}
 Arce, H. G., Borkin, M. A., Goodman, A. A., Pineda, J. E., \& Halle, M. W. 2010, \apj, 715, 1170 
\bibitem[Bessell \& Brett(1988)]{bessell1988}
 Bessell, M. S., \& Brett, J. M. 1988, \pasp, 100, 1134
\bibitem[Busquet et al.(2016)]{busquet2016}
 Busquet, G., Estalella, R., Palau, A., et al. 2016, \apj, 819, 139
\bibitem[Busquet et al.(2013)]{busquet2013}
 Busquet, G., Zhang, Q., Palau, A., et al. 2013, \apjl, 764, L26 
\bibitem[Chandrasekhar \& Fermi(1953)]{chandrasekhar1953}
 Chandrasekhar, S., \& Fermi, E. 1953, \apj, 118, 113 
\bibitem[Chapman et al.(2011)]{chapman2011}
  Chapman, N. L., Goldsmith, P. F., Pineda, J. L., Clemens, D. P., Li, D., \& Krco, M., 2011, \apj, 741, 21 
\bibitem[Clemens et al.(2012)]{clemens2012}
  Clemens, D. P., Pinnick, A., Pavel, M., \& Taylor, B. 2012, ApJS, 200, 19 
\bibitem[Churchwell et al.(2006)]{churchwell2006}
  Churchwell, E., Povich, M. S., Allen, D., et al. 2006, ApJ, 649, 759 
\bibitem[Crutcher et al.(2004)]{crutcher2004}
  Crutcher, R. M., Nutter, D. J., Ward-Thompson, D., \& Kirk, J. M., 2004, \apj, 600,  279
\bibitem[Davis (1951)]{davis1951}
  Davis, L. 1951, Phys. Rev., 81, 890 
\bibitem[Dobashi et al. (2019a)]{dobashi2019a}
 Dobashi, K., Shimoikura, T., Endo, N., Takagi, C., Nakamura, F.,
 Shimajiri, Y., \& Bernard, J.-Ph. 2019a, \pasj, in press 
\bibitem[Dobashi et al. (2019b)]{dobashi2019b}
 Dobashi, K., Shimoikura, T., Katakura, S., Nakamura, F., \& Shimajiri, Y. 2019b, \pasj, in press 
\bibitem[Elmegreen et al.(1979)]{elmegreen1979}
  Elmegreen, B. G., Lada, C. J., \& Dickinson, D. F. 1979, \apj, 230, 415 
\bibitem[Gaia Collaboration et al.(2018)]{gaia2018}
  Gaia Collaboration, Brown, A. G. A., Vallenari, A., et al. 2018, \aap, 616, 1 
\bibitem[Griffin et al.(2010)]{griffin2010}
  Griffin, M. J., Abergel, A., Abreu, A., et al. 2010, \aap, 518, L3 
\bibitem[Hanawa et al.(1992)]{hanawa1992}
  Hanawa, T., Nakamura, F., \& Nakano, T. 1992,\pasj, 44, 509 
\bibitem[Heiles(2000)]{heiles2000}
  Heiles, C. 2000, \aj, 119, 923  
\bibitem[Heiles \& Crutcher(2005)]{heiles2005}
  Heiles, C., \& Crutcher, R. 2005, in Lecture Notes in Physics Vol. 664, Cosmic Magnetic Fields, 
  ed. R. Wielebinski \& R. Beck (Berlin: Springer), 137 
\bibitem[Heyer et al.(1987)]{heyer1987}
  Heyer, M. H., Vrba, F. J., Snell, R. L., Schloerb, F. P., Strom, S. E., Goldsmith, P. F., \& Strom, K. M. 1987. \apj, 321, 855 
\bibitem[Hildebrand et al.(2009)]{hildebrand2009}
  Hildebrand, R. H., Kirby, L., Dotson, J. L., Houde, M., \& Vaillancourt, J. E. 2009, \apj, 696, 567 
\bibitem[Horiuchi et al.(1988)]{horiuchi1988}
  Horiuchi, T., Matsumoto, R., Hanawa, T., \& Shibata, K. 1988, \pasj, 40, 147
\bibitem[Ishii et al. (2019)]{ishii2019}
 Ishii, S., Nakamura, F., Shimajiri, Y., Kawabe, R., Tsukagoshi, T.,Dobashi, K., \& Shimoikura, T. 2019, \pasj, submitted
\bibitem[Ikeda \& Kitamura(2009)]{ikeda2009}
  Ikeda, N., \& Kitamura, Y. 2009, \apj, 705, 95
\bibitem[Inutsuka et al.(2015)]{inutsuka2015}
  Inutsuka, S., Inoue, T., Iwasaki, K., \& Hosokawa,T. 2015, \aap, 580, 49
\bibitem[Kandori et al.(2006)]{kandori2006}
 Kandori, R., Kusakabe, N., Tamura, M., et al.\ 2006, Proc. SPIE, 6269, 51 
\bibitem[Kirk et al.(2013)]{kirk2013}
 Kirk, H., Myers, P. C., Bourke, T. L.,  Gutermuth, R. A., Hedden, A., \& Wilson, G. W. 2013, \apj, 766, 115  
\bibitem[Konyves et al.(2010)]{konyves2010}
 Konyves, V.,Andre, P.,  Menshchikov, A. et al. 2010, \aap, 518, L106 
 \bibitem[Konyves et al.(2015)]{konyves2015}
  Konyves, V.,Andre, P.,  Menshchikov, A. et al. 2015, \aap, 584, A91  
\bibitem[Krumholz \& Tan(2007)]{krumholz2007}
 Krumholz, M. R., \& Tan, J. C. 2007, \apj, 654, 304 
\bibitem[Krumholz et al.(2014)]{krumholz2014}
 Krumholz, M. R., Bate, M. R., Arce, H. G., et al. 2014, In Protostars and Planets VI, 
 ed. H. Beuther, R. S. Klessen, C. P. Dullemond, \& T. Henning, 
 University of Arizona Press, Tucson, pp. 243-266. 
\bibitem[Kwon et al.(2015)]{kwon2015} 
 Kwon, J., Tamura, M., Hough, J. H., Nakajima, Y., Nishiyama, S., Kusakabe, N., Nagata, T., 
 \& Kandori, R. 2015, \apjs, 220, 17 
\bibitem[Kusune et al.(2015)]{kusune2015}
 Kusune. T., Sugitani, K.,  Miao, J., et al. 2015, \apj, 798, 60 
\bibitem[Kusune et al.(2016)]{kusune2016}
 Kusune, T., Sugitani, K., Nakamura, F., Watanabe, M., Tamura, M., Kwon, J., \& Sato, S. 2016, \apjl, 830, L23 
\bibitem[Kusune et al. (2019)]{kusune2019}
 Kusune, T., Nakamura, F., Sugitani, K., et al. 
 2019, \pasj, in press
 \bibitem[Lazarian \& Yuen(2018)]{lazarian2018}
 Lazarian, A.  \& Yuen, K. H. 2018, \apj, 853, 96 
\bibitem[Marchwinski et al.(2012)]{marchwinski2012}
 Marchwinski, R.C., Pavel, M. D., \& Clemens, D. P. 2012, \apj, 755, 130
\bibitem[Mathewson \& Ford(2000)]{mathewson1970}
 Mathewson, D. S., \& Ford, V. L. 1970, MmRAS, 74, 139
\bibitem[Matsumoto et al.(2015)]{matsumoto2015}
 Matsumoto, T., Dobashi, K., \& Shimoikura, T. 2015, ApJ, 801, 77 
\bibitem[McKee \& Ostriker(2007)]{mckee2007}
 McKee, C. F. \& Ostriker, E. C., 2007, \araa, 45, 56
\bibitem[Nagashima et al.(1999)]{nagashima99}
 Nagashima, C., Nagayama, T., Nakajima, Y., et al. 1999, in Star Formation 1999, ed. T. Nakamoto
(Nobeyama: Nobeyama Radio Observatory), 397
\bibitem[Nagayama et al.(2003)]{nagayama03}
 Nagayama, T., Nagashima, C., Nakajima, Y., et al. 2003, Proc. SPIE, 4841, 459 
\bibitem[Nakamura \& Li(2008)]{nakamura2008} 
 Nakamura, F., \& Li, Z.-Y. 2008, \apj, 687, 354 
\bibitem[Nakamura et al.(2011)]{nakamura2011}
 Nakamura, F., Sugitani, K., Shimajiri, Y., et al. 2011, \apj, 737, 56
\bibitem[Nakamura et al. 2014]{nakamura2014}
 Nakamura, F., Sugitani, K., Tanaka, T., et al. 2014, \apjl, 791, L23 
\bibitem[Nakamura et al. (2019a)]{nakamura2019a}
 Nakamura, F., Ishii, S., Dobashi, S., Shimoikura, T., Shimajiri, Y.,
 Kawabe, R., Tanabe, Y., Hirose, A. et al. 2019a, \pasj, in press  
 \bibitem[Nakamura et al. (2019b)]{nakamura2019b}
 Nakamura, F., Oyamada, S., Okumura, S., Ishii, S., Shimajiri, Y., Tanabe, Y., Tsukagoshi, T.,
 Kawabe, R., Momose, M. et al. 2019b, \pasj, in press 
\bibitem[Nakano \& Nakamura(1978)]{nakano1978}
 Nakano, T., \& Nakamura, T. 1978, \pasj, 30, 671 
\bibitem[Narayanan et al.(2012)]{narayanan2012}
 Narayanan, G., Snell, R., \& Bemis, A. 2012, MNRAS, 425, 2641 
\bibitem[Nishiyama et al.(2009)]{nishiyama2009}
 Nishiyama, S., Tamura, M., Hatano, H., Kato, D., Tanabe, T., Sugitani, \& K., Nagata, T. 2009, \apj, 696, 1407 
\bibitem[Nguyen- Luong et al.(2011)]{nguyen2011}
 Nguyen-Luong, Q., Motte, F., Hennemann, M., et al. 2011, \aap, 535, A76 
 \bibitem[Nguyen- Luong et al.(2016)]{nguyen2016}
Nguyen-Luong, Q., Nguyen, H. V. V., Motte, F. et al. 2016, \aap, 833, 23 
\bibitem[Nguyen-Luong et al.(2019)]{nguyen2019}
 Nguyen-Luong, Q.  et al. 2019, in preparation 
\bibitem[Ostriker et al. (2001)]{ostriker2001}
 Ostriker, E. C., Stone,J. M., \& Gammie, C. F. 2001, \apj, 546, 980 
\bibitem[Parker(1966)]{parker1966}
 Parker, E. N. 1966, \apj, 145, 811 
\bibitem[Planck Collaboration XXXV(2016)]{planckXXXV}
 Planck Collaboration XXXV, Ade, P. A. R., Aghanim, N., et al. 2016, \aap, 586, A138 
 \bibitem[Planck Collaboration XIX(2016)]{planckXIX}
 Planck Collaboration XIX, Ade, P. A. R., Aghanim, N., et al. 2016, \aap, 594, A19
\bibitem[Poglitsch et al.(2010)]{poglitsch2010}
 Poglitsch, A., Waelkens, C., Geis, N., et al. 2010, \aap, 518, L2 
\bibitem[Pattle et al.(2017)]{pattle2017}
 Pattle, K., Ward-Thompson, D., Kirk, J. M., et al. 2017, \mnras, 464, 4255
\bibitem[Povich et al.(2016)]{povich2016} 
 Povich, M. S., Townsley, L. K., Robitaille, T. P., Broos, P. S., Orbin, W. T., King, R. R., Naylor,T., 
 \& Whitney, B. 2016, \apj, 825, 125
\bibitem[Povich \& Whitney(2010)]{povich2010}
 Povich, M. S., \& Whitney, B. A. 2010, \apjl, 714, L285  
\bibitem[Santos et al.(2016)]{santos2016}
 Santos, F. P., Busquet, G., Franco, G. A. P., Girart, J. M., \& Zhang, Q. 2016, \apj, 832, 186 
\bibitem[Sato et al.(2010)]{sato2010}
 Sato, M., Hirota, T., Reid, M. J., et al. 2010, \pasj, 62, 287 
\bibitem[Shimajiri et al.(2014)]{shimajiri2014}
 Shimajiri, Y., Kitamura, Y., Saito, M. 2014, \aap, 564, A68 
\bibitem[Shimoikura et al.(2015)]{shimoikura2015}
 Shimoikura, T., Dobashi, K., Nakamura, F., Hara, C., Tanaka, T., Shimajiri, Y., Sugitani, K., \& Kawabe, R.
2015, \apj, 806, 201
\bibitem[Shimoikura et al.(2019a)]{shimoikura2019a}
 Shimoikura, T., Dobashi, K., Nakamiura, F., Shimojiri, Y., \& Sugitani, K. 2019a, PASJ, in press (arXiv:1809.09855)
\bibitem[Shimoikura et al. (2019b)]{shimoikura2019b}
 Shimoikura, T., Dobashi, K., Hirose, A., Nakamura, F., Shimajiri, Y.,  \& Sugitani, K. 2019b, \pasj, in press 
\bibitem[Shu et al.(1987)]{shu1987}
 Shu, F. H., Adams, F. C., \& Lizano, S. 1987, \araa, 25, 23
\bibitem[Skrutskie et al.(2006)]{skrutskie06}
 Skrutskie, M. F., et al. 2006, \aj, 131, 1163
\bibitem[Soler \& Hennebelle(2017)]{soler2017b}
 Soler, J. D., \& Hennebelle, P. 2017, \aap, 607, 2
\bibitem[Soler et al.(2017)]{soler2017a}
 Soler, J. D., Ade, P. A. R., Angile, F. E., et al. 2017, \aap, 603, A64 
\bibitem[Soler et al.(2013)]{soler2013}
 Soler, J. D., Hennebelle, P., Martin, P. G., et al. 2013, \apj, 774, 128
\bibitem[Sugitani et al.(2011)]{sugitani2011}
 Sugitani,K., Nakamura, F., Watanabe, M. et al. 2011, \apj, 734, 63 
\bibitem[Tackenberg et al.(2014)]{tackenberg2014}
 Tackenberg, J., Beuther,H., Henning, Th., Linz, H., Sakai, T., Ragan, S. E., Krause, O., Nielbock, M., 
 Hennemann, M., Pitann, J. \& Schmiedeke, A. 2014, \aap, 565, A101 
\bibitem[Tanabe et al. (2019)]{tanabe2019}
 Tanabe, Y.,  Nakamura, F., Tsukagoshi, T., Shimajiri, Y.,Sasaki, K., Ishii, S., Kawabe, R., 
 Feddersen, J. et al. 2019, \pasj, submitted
\bibitem[Tanaka et al.(2013)]{tanaka2013}
 Tanaka, T., Nakamura, F., Awazu, Y., et al. 
 2013, \apj, 778, 34
\bibitem[Tamura et al.(1987)]{tamura1987}
 Tamura, M., Nagata, T., \& Sato, S. 1987, \mnras, 224, 413 
\bibitem[Tatematsu et al.(1990)]{tatematsu1990}
 Tatematsu, K., Fukui, Y., Landecker,T. L., \& Roger, R. S. 1990, \aap, 237, 189 
\bibitem[Urquhart et al.(2007)]{urquhart2007}
 Urquhart, J. S., Busfield, A. L., Hoare, M.G., Lumsden, S. L., Clarke, A. J., Moore, T. J. T., 
 Mottram, J. C., \& Oudmaijer, R. D., 2007, \aap, 461, 11
\bibitem[Vrba et al.(1976)]{vrba1976}
 Vrba, F. J., Strom, S. E., \& Strom, K. M. 1976, \aj, 81, 958 
\bibitem[Wardle \& Kronberg(1974)]{wardle74} 
 Wardle, J. E. C., \& Kronberg, P. P. 1974, \apj, 194, 249 
\bibitem[Wainscoat(1992)]{wainscoat1992}
 Wainscoat, R. J., Cohen, M., Volk, K., et al. 1992, ApJS, 83, 111
\bibitem[Wilking et al.(1979)]{wilking1979}
 Wilking, B. A., Lebofsky, M. J., Rieke, G. H., \& Kemp, J. C. 1979, \aj, 84, 199 
\bibitem[Ward-Thompson et al.(2016)]{ward-thompson2016}
 Ward-Thompson, D., Pattle, K., Kirk, J. M., et al., 2016, \mnras, 463, 1008
\bibitem[Zuckerman \& Evans(1974)]{zuckerman1974}
 Zuckerman, B. \& Evans, N. J., II, 1974, \apj, 192, 149
\end{thebibliography}
\end{document}